\documentclass[useAMS,usenatbib,usegraphicx]{mn2e}
\usepackage{float}
\usepackage{epsf,rotating,url}
\usepackage{subfigure,amssymb,amsfonts,amstext,amsgen,amsopn,amsxtra,indentfirst,lscape,times,rotating}
\usepackage{longtable}
\usepackage{graphics}
\usepackage{keyval}
\usepackage{soul}
\usepackage{trig}
\usepackage{dcolumn}
\usepackage{bm}
\def\beq{\begin{equation}}
\def\eeq{\end{equation}}
\def\bey{\begin{eqnarray}}
\def\eey{\end{eqnarray}}
\def\msun{M_\odot}
\def\lsun{L_\odot}
\def\kms{\, {\rm km \, s}^{-1} }

\def\grad{{\vec \nabla}}

\def\mnras{MNRAS}
\def\apj{ApJ}
\def\nat{Nature}

\def\apjl{ApJ}
\def\na{New Astronomy}
\def\araa{ARAA}

\def\aap{A \& A}
\def\aj{AJ}

\def\aap{Astron. Astrophys.}

\title{N-body simulations of the Carina dSph in MOND}

\author[G. W. Angus, G. Gentile, A. Diaferio,  B. Famaey, \\ K. J. van der Heyden]{G. W. Angus$^{1,2}$\thanks{E-mail: garry.angus@vub.ac.be}, G. Gentile$^{3,1}$, A. Diaferio$^{4,5}$, B. Famaey$^{6}$, and K. J. van der Heyden$^{2}$ \\ 
$^{1}$Department of Physics and Astrophysics, Vrije Universiteit Brussel, Pleinlaan 2, 1050 Brussels, Belgium \\
$^{2}$Astrophysics, Cosmology \& Gravity Centre, Dept. of Astronomy, University of Cape Town, Private Bag X3, Rondebosch, 7701, South Africa \\
$^{3}$Sterrenkundig Observatorium, Universiteit Gent, Krijgslaan 281, 9000, Gent, Belgium\\
$^{4}$Dipartimento di Fisica, Universit\`a di Torino, Via P. Giuria 1, I-10125, Torino, Italy \\
$^{5}$Istituto Nazionale di Fisica Nucleare, Via P. Giuria 1, I-10125, Torino, Italy\\
$^{6}$Observatoire Astronomique de Strasbourg, CNRS UMR 7550, France \\
}
\begin{document}

\date{\today}
\maketitle
\begin{abstract}
The classical dwarf spheroidals (dSphs) provide a critical test for Modified Newtonian Dynamics (MOND) because they are observable satellite galactic systems with low internal accelerations and low, but periodically varying, external acceleration. This varying external gravitational field is not commonly found acting on systems with low internal acceleration. Using Jeans modelling, Carina in particular has been demonstrated to require a V-band mass-to-light ratio greater than 5, which is the nominal upper limit for an ancient stellar population. We run MOND N-body simulations of a Carina-like dSph orbiting the Milky Way to test if dSphs in MOND are stable to tidal forces over the Hubble time and if those same tidal forces artificially inflate their velocity dispersions and therefore their apparent mass-to-light ratio. We run many simulations with various initial total masses for Carina, and Galactocentric orbits (consistent with proper motions), and compare the simulation line of sight velocity dispersions (losVDs) with the observed losVDs of Walker et al. (2007). We find that the dSphs are stable, but that the tidal forces are not conducive to artificially inflating the losVDs. Furthermore, the range of mass-to-light ratios that best reproduces the observed line of sight velocity dispersions of Carina is 5.3 to 5.7 and circular orbits are preferred to plunging orbits. Therefore, some tension still exists between the required mass-to-light ratio for the Carina dSph in MOND and those expected from stellar population synthesis models. It remains to be seen whether a careful treatment of the binary population or triaxiality might reduce this tension.
\end{abstract}

\section{Introduction}
\protect\label{sec:intr}
The classical dwarf spheroidal galaxies of the Milky Way are eight low surface brightness galaxies that are currently at distances between 60 and 250~kpc. They have total luminosities in the V-band ranging from $L_V\sim4\times10^5$ to $1.7\times10^7\lsun$ (\citealt{mateo98}) and sizes of order a kiloparsec. For comparison, the Milky Way luminosity and size are $L_V\sim 6\times10^{10}\lsun$ (\citealt{mcgaugh08}) and $\sim30~kpc$. Clearly, such puny luminosities within relatively large volumes earns the dwarf spheroidals (dSphs) their low surface brightness moniker and also puts them in a very interesting category since low surface brightness galaxies typically have large dark matter (DM) components.

Being spheroidal systems, information about their dynamical mass can be obtained from Jeans modelling of their stellar velocity dispersions (see \citealt{mamon10} for more information). For this reason, \cite{walker07} obtained hundreds of spectra of probable member stars for each of the dSphs, sampled over their full projected areas. Photometrically and spectroscopically identified interloper stars (non members, typically foreground stars) were rejected and each dSph's projected velocity dispersion, as a function of projected radius, was computed. They then performed Jeans modelling of each dSph, which employs the observed stellar surface brightness profile and fits for the unknown DM profile, by comparing modelled with observed projected velocity dispersions. This blatantly showed that the dSphs are some of the most DM dominated (in Newtonian dynamics) galaxies in the Universe.

Although the dynamics of the dSphs can be easily explained by the presence of DM, there are other peculiarities related to their phase-space distribution around the Milky Way which makes one question this conclusion. The major open questions relating to dSphs are comprehensively reviewed in \cite{walker14}, but we restate them here. First of all, from comparison with cold dark matter (CDM) only cosmological simulations (like those of \citealt{klypin99,moore99}) one would naively expect a greater number of these satellite galaxies within 250~kpc of the Milky Way. Certain authors like \cite{benson02,munoz09,maccio10,lihelmi10} have suggested that this lack of satellites may be due to star formation inefficiencies due to re-ionisation and supernova feedback in these lower mass CDM halos which only enables a fraction of all halos to form stars. However, this fails to address the problem noted by \cite{boylan12} that associating the dSphs with the most massive Milky Way subhalos, as we expect in these models, is incompatible with the relatively low masses and densities of the measured DM halos.

The other more pressing concern is that the dSphs are not isotropically distributed around the Milky Way. Rather, they are distributed as a great rotationally-supported disk that is surprisingly thin, with an RMS thickness of 10-30~kpc (see \citealt{metz08} and the detailed review of \citealt{kroupa10}), which is substantially smaller than typical RMS thicknesses in nearby groups of galaxies. If it were an isolated incident, this would be less troubling, but \cite{ibata13} have recently shown a similar structure in the satellite galaxy distribution surrounding the M31 galaxy with an RMS thickness of less than 14.1~kpc (with 99\% confidence) to which half the satellites belong. Furthermore, \cite{chiboucas13} have recently identified a flattened distribution of satellites around M81.

These satellite distributions have been shown to be highly unlikely to arise from CDM cosmological simulations, although once in place they could naturally be stable (\citealt{adk11,pawlowski12,deason11,bowden13}).

On the other hand, following a merger or a flyby (e.g., \citealt{zhao13}) between two galaxies, with mass ratios between 1:1 and 1:4, the probability of forming such a polar disk of satellites could easily reach 50\% (\citealt{pawlowski12}).

Separately, there are observations of dwarf galaxies forming out of the tidal debris produced from a wet galactic merger (\citealt{bournaud07}), which may demonstrate evidence for MOND (\citealt{gentile07a,milgrom07b}).

Returning to the dSphs, if they are in fact tidally formed they should not have large DM abundances. Furthermore, they have very little neutral hydrogen (\citealt{mateo98}) and no significant emission from molecular gas. However, these eight classical dSphs do require large DM abundances when interpreted with Newtonian dynamics, and they have a peculiar orbital distribution that may be difficult to explain within the CDM framework. Therefore, it is worth investigating their dynamics in an alternative theory of gravity that can, in principal, be consistent with the merger scenario and the large velocity dispersions without galactic DM. One such alternative is Modified Newtonian Dynamics (MOND; \citealt{milgrom83a} and see \citealt{famaey12} for a thorough review).

\cite{brada00b} used a particle-mesh N-body solver to study the influence of the Milky Way on the dSphs. Their work preceded the high quality velocity dispersion data, but demonstrated that there are orbital regions where dSphs can orbit with adiabatic (reversible) changes to their velocity dispersion and density profiles. In addition, there are non-adiabatic regions where the rapid change of the external gravitational field of the Milky Way disturbs the density profile at pericentre and this does not recover by the time the dSph returns to apocentre. Finally, there are tidal regions where mass will be stripped from the dSphs at pericentre.

Using the data of \cite{walker07}: \cite{angus08} and \cite{serra10}  performed Jeans modelling in MOND. There, the goal was to isolate the two free parameters: the mass-to-light ratio of the stellar population and the velocity anisotropy. Velocity anisotropy is the {\it a priori} unknown relationship between the probability of radial and tangential stellar orbits within the dSph. This can also be used as a free parameter in the context of DM halo fitting, but is somewhat redundant given the freedom of possible DM halo profiles. In MOND, it is an essential ingredient to alter the shape of the projected velocity dispersion profile, whereas all the mass-to-light ratio can do is raise or lower the amplitude of the velocity dispersions. \cite{angus08} found that the four dSphs with the highest surface brightness (highest internal gravities) had reasonable mass-to-light ratios, but the other four required mass-to-light ratios that were larger than the expected range of 1 to 5 in the V-band found from stellar population modelling (\citealt{maraston05}).

Much simulation work has been done in this vein in the standard paradigm (see e.g. \citealt{kroupa97,klessen03,read06,penarrubia09,klimentowski09}). More specifically, the work of \cite{munoz08} focused on a very similar thesis as ours, which was whether tidally disturbed mass-follows-light models of a DM dominated Carina dSph are consistent with the observed projected surface density and projected velocity dispersion profile. Those authors found that there were indeed combinations of mass and orbital parameters that could faithfully reproduce the Carina dSph.

\cite{sanchez07} investigated the likelihood of survival for the dSphs in MOND after successive orbits over a Hubble time. They found that only Sextans was likely to dissolve in less than a few Gyr, but that the deduced dynamical mass-to-light ratios of Ursa Minor and Draco (out of the eight classical dSphs) were too large to be consistent with only the stellar populations. They also showed, based on their current positions, that tidal stirring might be an important consideration for Sextans, Sculptor and Ursa Minor, but not Carina. Other relevant work was carried out by \cite{sanchez10} and \cite{lora13} who looked at the importance of cold kinematic substructures that are found in the Sextans and Ursa Minor dSphs. It was shown their longevity can be used to discriminate between modified gravity and CDM.

Given the separation in surface brightness between dSphs that satisfied MOND and those that did not, it was suggested in \cite{angus08} that the latter four dSphs may be subject to tidal forces that produce tidally unbound interloper stars and inflate the velocities of the bound stars.

Our aim here is to test this hypothesis by running high resolution MOND N-body simulations of satellite galaxies orbiting the Milky Way and comparing the simulated projected velocity dispersions with the observed ones. Insodoing we also hope to elucidate the zones of possible orbits open to the satellites without being torn to shreds by the Milky Way. This is an essential sanity check for when high accuracy proper motions become available.

We focus on the Carina dSph because out of the four least luminous classical dSphs it has a well measured surface brightness profile, large numbers of stellar line of sight velocities for Jeans modelling and relatively accurately measured proper motions.

In Section 2 we present the Jeans analysis, in Section 3 we discuss how to incorporate the external field and the setup of our simulations. In Section 4 we compare simulated with observed projected velocity dispersions, in Section 5 we give our results, and finally in Section 6 we draw our conclusions.

\section{Jeans analysis}
\subsection{Likelihood analysis of the Carina dSph's observed projected velocity dispersion profile}
\protect\label{sec:jeans}

Dynamical interloper stars, i.e. those not identified photometrically or spectroscopically, can be dealt with using various algorithms (none of which is fully accurate). At least two such methods have been applied to the dSphs, \cite{klimentowski07} and \cite{serra10}, and removed significant numbers of presumed interlopers. These interlopers are typically tidally stripped stars that are mostly removed during pericentre. In \S\ref{sec:compar}, we will be comparing our simulation velocity dispersions with the observed ones. This observational data is expected to contain dynamical interlopers, and our simulations - if valid representations of the real dSphs - should therefore produce comparable numbers of interlopers (assuming they originate from the dSph and not the Milky Way). 

For the above reason, we re-bin data from \cite{walker07} without any removal of dynamical interlopers beyond that performed by those authors. We separated the data into projected radius bins of 50~pc with unequal numbers of stars per bin. The projected velocity dispersion and its associated uncertainty, in each bin, is calculated by making a Markov Chain Monte Carlo (MCMC) analysis using Eq~B1 of \cite{hargreaves94}
\beq
P(v_i)={1 \over \sqrt{2\pi(\sigma_i^2+\sigma_v^2)}}exp\left[{-v_i^2 \over 2(\sigma_i^2+\sigma_v^2)} \right].
\eeq

This equation uses the different measurement errors of each individual star to weight their contribution to the line of sight velocity dispersion in each radius bin. It gives the probability of a sample of stars with zero systemic velocity, with projected velocity, $v_i$, and velocity uncertainty, $\sigma_i$, being chosen from a distribution with velocity dispersion, $\sigma_v$.

The likelihood is formed by the product of these probabilities for the number of stars in that bin. The MCMC analysis fits for the velocity dispersion and uncertainty in each bin by sampling the likelihood and producing a cumulative likelihood distribution. This allows us to ascertain the maximum likelihood velocity dispersion and 1$\sigma$ error for each radius bin.

\begin{figure}
\includegraphics[angle=0,width=8.0cm]{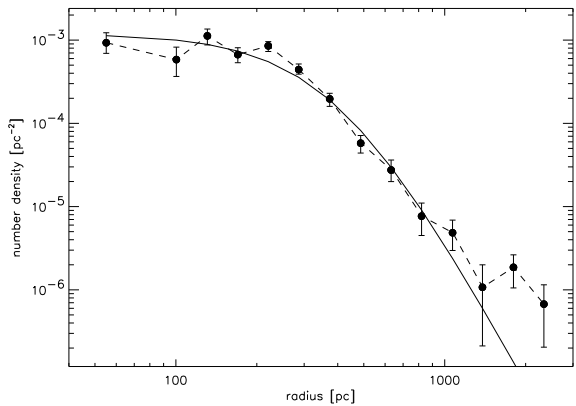}
\caption{Here we plot the projected number density of stars in Carina taken from Munoz et al. (2006). The solid line is the re-normalised projection of our best fit 3D stellar density model with $\rho_*(r)=M/L_V \times3.8\times10^{-3}\left[1+(r/410 \rm pc)^2 \right]^{-3.5} \msun pc^{-3}$.}
\label{fig:munoz}
\end{figure}
\subsection{Approximate Jeans Equation for MOND}
\protect\label{sec:jeansmond}
The Jeans equation for a spherical galaxy solves for the radial velocity dispersion, $\sigma_r$, requiring knowledge of the logarithmic density slope $\alpha(r)={d\ln\rho_* \over d\ln r}$ of the tracers (stars in the dSph's case), velocity anisotropy $\beta=1-{\sigma^2_t \over 2\sigma^2_r}$ - where $\sigma_t$ is the tangential velocity dispersion. It also assumes knowledge of the gravity profile, $g(r)$, which is usually based on fitting for the unknown parameters of the DM halo, or the mass-to-light ratio in non DM models. This gives

\beq
\protect\label{eqn:jeans}
{d \over dr}\sigma_r^2(r) + {\alpha(r)+2\beta \over r}\sigma_r^2(r) = -g(r)
\eeq

Using geometrical arguments, it is possible to convert the radial velocity dispersion into a line of sight (or projected) velocity dispersion, which is used to compare with the observed velocity dispersion.
\subsection{Likelihood analysis of the Carina dSph's mass-to-light ratio and velocity anisotropy in MOND}
\protect\label{sec:likejeans}
In order to re-emphasise the likelihood of the two free parameters in the MOND Jeans analysis of the Carina dSph galaxy, we performed another MCMC analysis. The modelled, projected velocity dispersion is a function of the surface brightness profile, stellar mass-to-light ratio, stellar velocity anisotropy, Galactocentric distance, and Milky Way mass (see \S\ref{sec:jeansmond} and Eqs~\ref{eqn:jeans} and \ref{eqn:efe} for the Jeans equation relating these parameters). We consider all parameters fixed except the constant (with radius) velocity anisotropy and mass-to-light ratio. These two we allow to vary, and we produce likelihood plots after comparing the modelled projected velocity dispersion with the observed one.

\begin{figure*}
\centering
\subfigure{
\includegraphics[angle=0,width=8.0cm]{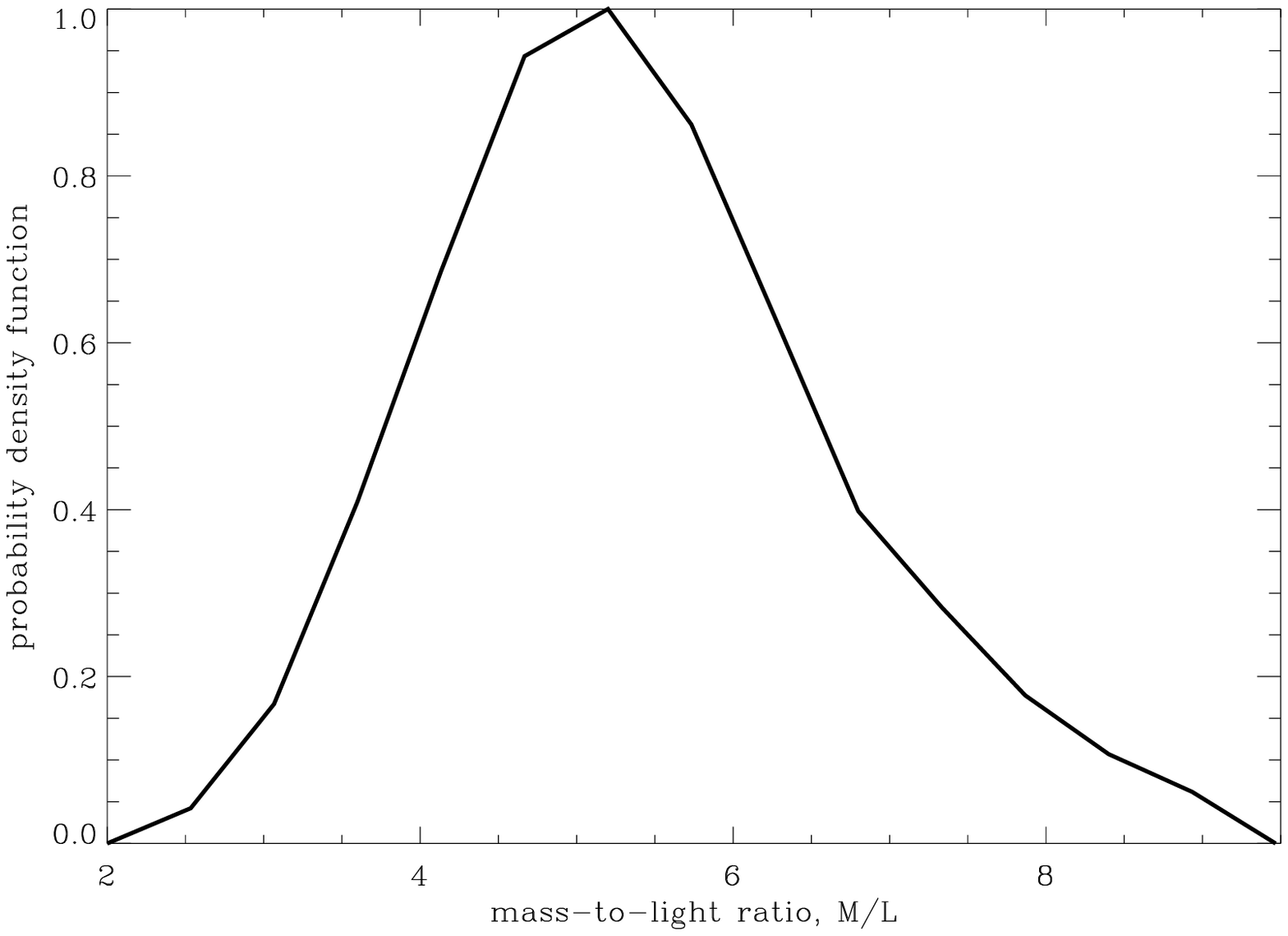}
}
\subfigure{
\includegraphics[angle=0,width=8.0cm]{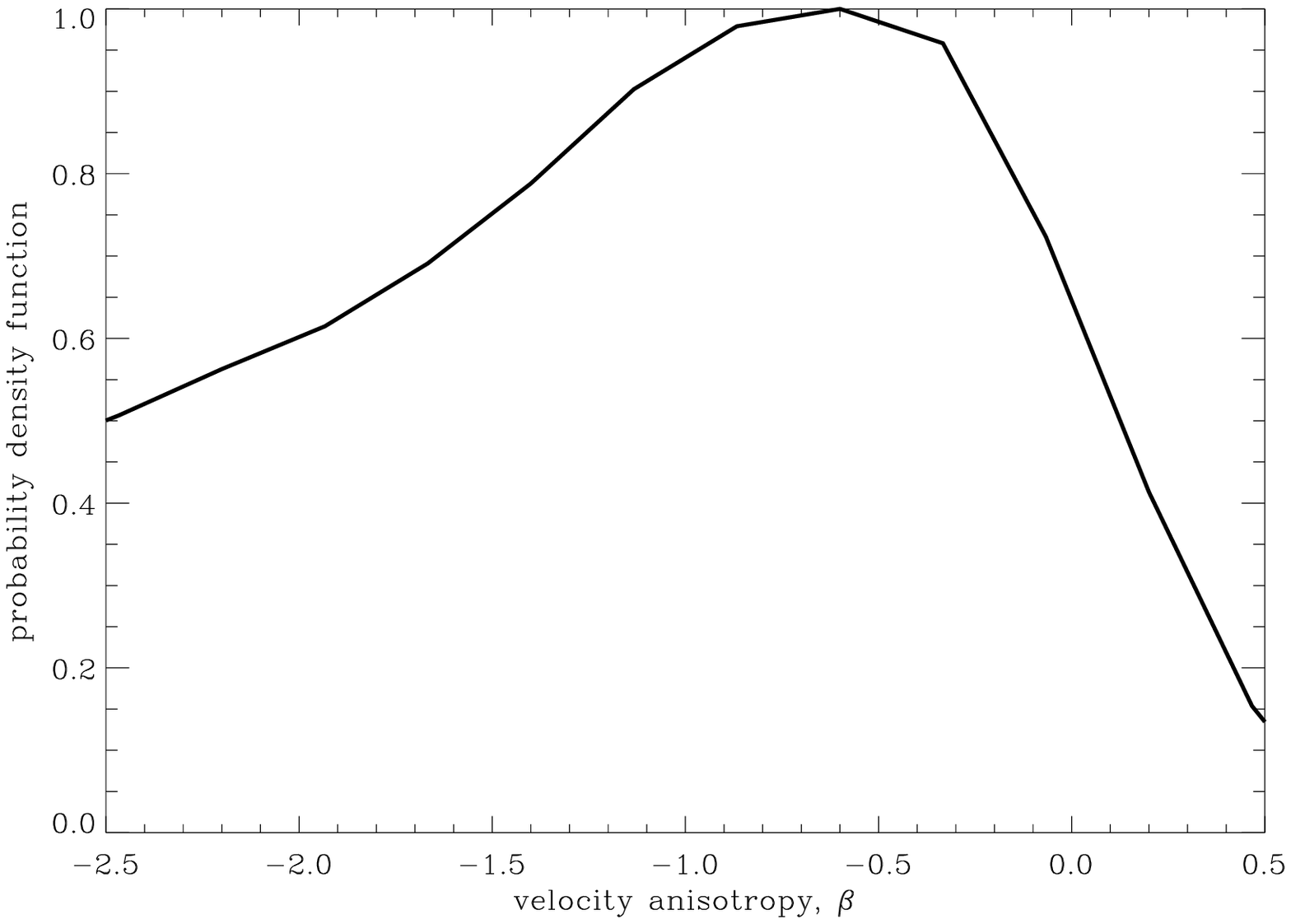}
}\\
\caption{Here we plot the normalised probability distributions of the MOND stellar mass-to-light ratio (left hand panel) and velocity anisotropy, $\beta$, (right hand panel) from a Jeans modelling MCMC analysis of the projected velocity dispersions measured by Walker et al. (2007).}
\label{fig:m2l_pdf}
\end{figure*}

We have fixed the baryonic mass of the Milky Way to be $M_{MW}=6\times10^{10}\msun$ (\citealt{mcgaugh08}) and use a Milky Way-Carina distance of $r_{MW}=101~kpc$ (\citealt{mateo98}). We use a 3D light distribution of the form $\rho_*(r)=M/L_V\times \rho_{*,0}(1+(r/r_c)^2)^{-\alpha}$ which we fitted to the observed surface brightness profile of Carina from \cite{munoz06} in Fig~\ref{fig:munoz}. We use $\rho_{*,0}=3.8\times10^{-3}\msun pc^{-3}$, $r_c=410~pc$ and $\alpha=3.5$ to give an $M/L_V=1$ total luminosity of $L_V=4.4\times10^5\lsun$ (\citealt{mateo98}; for which the uncertainty is $\sim$27\%). We also use Eq~\ref{eqn:efe} to include the external field effect of MOND, the appropriateness of which we discuss in \S\ref{sec:efe}. 

In the left hand panel of Fig~\ref{fig:m2l_pdf} we plot the probability distribution functions for Carina's mass-to-light ratio and velocity anisotropy using MCMC Jeans modelling of the projected velocity dispersion profile. One can see there is a strong preference for a mass-to-light ratio larger than 3 and the maximum likelihood with 1$\sigma$ uncertainty is 5.2$\pm$1.2. The most probable velocity anisotropy is $\beta=-0.8$ (see right hand panel of Fig~\ref{fig:m2l_pdf}) and isotropic (as well as radial) orbits are disfavoured. If we add Carina's distance as a free parameter, with a prior set by its observational error, we find no significant change in the maximum likelihood solution for the mass-to-light ratio. However, the 1$\sigma$ confidence limits increase by 15\%. According to \cite{maraston05}, the upper limit for a mass-to-light ratio in the V-band for an old population of stars is 5. Therefore, the modelled mass-to-light ratio is at the high end of the expected range. Furthermore, a mass-to-light ratio of 5 requires a particular initial mass function and Carina is not formed purely by an ancient stellar population. In fact, \cite{mateo98} shows there are several star forming epochs, with one strong burst around $6\pm1$~Gyr ago. Additionally, Ursa Minor, Draco and Sextans appear to require higher than expected mass-to-light ratios.

\section{Incorporating the external field}
\protect\label{sec:efe}

MOND is an alternative theory of gravity which removes the need for galactic DM by appealing to stronger than Newtonian gravity in regions of weak acceleration. Its phenomenological basis lies in the constancy of the outer parts of rotation curves of spiral galaxies, the baryonic Tully-Fisher relation and the apparent one-to-one correspondence between baryonic density and the observed dynamics of the large majority of galactic systems. This last point means that knowledge of the baryonic matter distribution gives full knowledge of the DM distribution when Newtonian dynamics are used. In MOND, an isolated spherical galaxy with Newtonian internal gravity $g_{n,i}=GM(r)r^{-2}$ will actually produce a gravitational field according to $g_{i}=\nu(|g_{n,i}|/a_o)g_{n_i}$, where $\nu$ is an interpolating function that allows the gravitational field to transition from the Newtonian dynamics we experience in the Solar System, or the bright nuclei of galaxies, to the necessarily amplified gravity at the edges of spiral galaxies.

In this work we have chosen the interpolating function 
\beq
\protect\label{eqn:nu}
\nu(y)=0.5+0.5\sqrt{1+4/y},
\eeq
(see \citealt{fb05} and \citealt{famaey12} for a discussion). This factor $\nu$ is therefore greater than or equal to unity and it will become larger the weaker the internal gravity of the system is. This means that the central regions of dense elliptical galaxies will have $\nu\approx1$, Milky Way like spiral galaxies will have $\nu$ of around 1.5-2 near the Sun's position and rising thereafter. Finally, galaxies with very low stellar densities will have large $\nu$, where $\nu \rightarrow \sqrt{a_o/|g_{n,i}|}$ which leads directly to flat rotation curves and the baryonic Tully-Fisher relation.

There is of course one final complication, which is the result of MOND breaking the strong equivalence principle (\citealt{milgrom86}). In MOND, the internal gravity of a satellite galaxy of the Milky Way is determined not only by the satellite's stellar distribution (and therefore mass distribution), but also by the local strength of the Milky Way's gravitational field. This effect should not be confused with tidal forces, which also exist.

\begin{figure}
\includegraphics[angle=0,width=8.0cm]{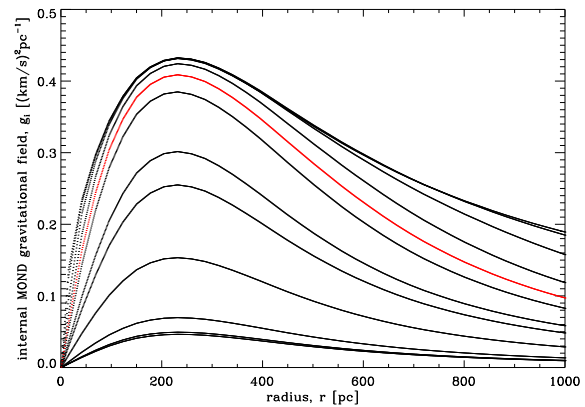}
\caption{Here we plot the internal gravity profile of Carina using various strengths for the external field. We use the parameters presented in \S\ref{sec:likejeans}. The external field is handled using the boundary conditions method described in \S\ref{sec:constantefe} (not the two-component simulations). Starting with the top curve and going to the lowest, the strengths of the external field in units of $a_o$ are $10^{-4}$, $10^{-3}$, $2.5\times 10^{-3}$, $5\times 10^{-3}$, $7.367\times 10^{-3}$ (red), 0.01, 0.02, 0.03, 0.1, 1, 10 and 100. The upper curve is the MOND limit and the bottom curve is the Newtonian limit. The red curve is the gravity profile for the external field Carina is currently experiencing.}
\label{fig:efe}
\end{figure}

The tidal force imposes a force gradient across the satellite from the far to near side (relative to the Milky Way). But in MOND, if the magnitude of the Milky Way's gravitational field is constant across the satellite, the satellite's internal gravity will still be altered by that constant field. This phenomenon does not exist in Newtonian dynamics. Increasing the magnitude of the Milky Way's gravitational field, by locating the satellite closer to it, will reduce the internal gravitational field of the satellite towards the Newtonian limit i.e. $\nu \rightarrow$ constant. Decreasing the magnitude of the Milky Way's gravitational field, by moving the satellite  further from it, will increase the internal gravitational field of the satellite towards the MONDian limit. 
The crucial qualitative corollaries are explored next and can be gleaned from a comparison with a satellite galaxy in Newtonian dynamics.

In Newtonian dynamics a satellite galaxy must have a DM halo that outweighs the stars by a factor between ten and a hundred. It is this DM halo that provides the supplementary force to bind the stars, without which the satellite would dissolve in a few dynamical times. Furthermore, the Milky Way has a DM halo which provides the enhanced gravity, over the stars, to bind the satellite galaxy to it.

In MOND, the satellite galaxy has no DM halo, but the boosted gravity of MOND provides the supplementary force to bind the stars. The Milky Way also has no DM halo, but again MOND - sourced by the Milky Way's baryons - provides the supplementary gravity to bind the satellite to the Milky Way.

To illustrate the difference between the dSphs in MOND and Newtonian dynamics, let's say the satellite makes an orbit from apocentre to pericentre. In Newtonian Dynamics, the satellite starts the orbit at apocentre with a particular internal gravitational field that drives the large velocity dispersions and at pericentre this gravitational field is virtually unchanged since the DM halo distribution has barely changed. The only difference at pericentre is that tidal forces from the Milky Way start to influence it.

In MOND, the satellite starts the orbit at apocentre with strong internal gravity, but as it approaches pericentre the internal gravity becomes progressively weaker because of the increasing external gravity of the Milky Way, which diminishes the internal gravity of the satellite. In addition, the tides get stronger the closer the dSph gets to pericentre. Therefore, precisely at the moment when the satellite requires additional gravity to protect it from the tidal forces, it loses it, making observations of tidal harassment of the Carina dSph by \cite{battaglia12} a natural expectation. This makes satellite galaxies in MOND far more susceptible to tidal destruction than those in Newtonian dynamics with DM. As such, the dearth of satellite galaxies near to the Milky Way might be no surprise in MOND, but many more satellites are expected in CDM.

In the remainder of this section we discuss various methods for including the external field of a host galaxy acting upon a satellite galaxy in MOND. There are three main ideas we introduce, the first two being only approximations. The first one is the simple case where the external field only enters the $\nu$ function (Eq~\ref{eqn:nu}). This approximation of the external field can be defined by Eq~\ref{eqn:efe} and we only use this in the Jeans analysis of \S\ref{sec:likejeans} and to compare the gravitational field of a test satellite with the other methods - it is not used in any simulations.

The second method is to include the external field through the boundary conditions of the particle-mesh Poisson solver. This is described in \S\ref{sec:constantefe} but, as per the first case, this method is only used to make comparisons of the gravitational field of a test satellite with the other two methods.

The final method is the one we use in all the simulations. We call this two-component simulations because we not only simulate particles representing the satellite (dSph - Carina), but also we simulate a coarse representation of the host (Milky Way). These two component simulations can account for both the external field and tidal forces and we describe it in more detail in \S\ref{sec:2comp}.

We compare the gravitational fields produced by these three methods in \S\ref{sec:efp}.

\subsection{Approximated MOND external field prescriptions for satellite galaxies}
\protect\label{sec:mond4sat}

In this sub-section we present the well-known equations that govern the incorporation of the external field in MOND. The field equation to solve for the MOND potential (\citealt{milgrom10}), $\Phi$ is
\beq
\protect\label{eqn:qumond2}
\grad \cdot (\grad\Phi)=\grad \cdot \left[ \nu(|\grad\Phi_n| / a_o) \grad\Phi_n \right],
\eeq
$a_o$ is the MOND acceleration constant chosen here to be $3.6~(\kms)^2pc^{-1}$. $\Phi_n$ is the Newtonian potential which is solved from the matter density, $\rho$, using the Poisson equation $\nabla^2\Phi_n=4\pi G \rho$.

If we are considering a satellite in orbit of a host galaxy, we can separate $\grad\Phi$ into an internal, $\grad\Phi_i$, and an external, $\grad\Phi_e$, gravitational field. This gives, after removing the divergences and ignoring the curl-field:

\begin{multline}
\protect\label{eqn:efepar}
\grad\Phi_i=\nu\left({ |\grad\Phi_{n,i}+ \grad\Phi_{n,e}| \over a_o}\right) \grad\Phi_{n,i}\\+\left(\nu\left({| \grad\Phi_{n,i}+ \grad\Phi_{n,e}| \over a_o}\right)-\nu\left({ |\grad\Phi_{n,e} |\over a_o}\right) \right)\grad\Phi_{n,e}.
\end{multline}
If we now crudely consider only directions in the plane perpendicular to the external field we can ignore the second term in Eq~\ref{eqn:efepar} because the external gravitational field's magnitude in that direction is zero. However, in the argument of the $\nu$ function for the first term of Eq~\ref{eqn:efepar} we must include the modulus of all gravitational fields regardless of direction. In the direction perpendicular to the external field we can add the external field and the internal field in quadrature in the argument of the $\nu$ function, which gives

\beq
\protect\label{eqn:efe}
\grad\Phi_i=\nu\left({ \sqrt{|(\grad\Phi_{n,i})^2+ (\grad\Phi_{n,e})^2|} \over a_o}\right) \grad\Phi_{n,i}.
\eeq
Thus, in the limit $\grad\Phi_{n,i}>>\grad\Phi_{n,e}$, we can ignore the external gravitational field, $\grad\Phi_{n,e}$ and we reduce Eq~\ref{eqn:efepar} (or Eq~\ref{eqn:efe}) to the standard MOND formula in spherical symmetry, $\grad\Phi_i=\nu\left({| \grad\Phi_{n,i} |\over a_o}\right) \grad\Phi_{n,i}$.
If on the other hand $\grad\Phi_{n,e}>>\grad\Phi_{n,i}$ everywhere, then the gravitational field of the satellite is simply a scaled up version of the Newtonian internal gravitational field i.e. it is as if Newton's gravitational constant has been revised to a value $G\times \nu\left({ |\grad\Phi_{n,e}| \over a_o}\right)$. As with most things, the interesting cases lie in the middle ground, so we typically use Eq~\ref{eqn:efe} if we perform a curl-free analysis including the external field effect.
\subsection{Constant external field}
\protect\label{sec:constantefe}
On the topic of handling the external field numerically, we modified our openMP parallelised QUMOND galactic Poisson solver code, that was introduced in \cite{angus12} to fit the rotation curves of a sample of spiral galaxies, to account for the external field. The code uses a refinement strategy to go from the coarsest grid to the finest grid, each time halving the size of the box and thus doubling the spatial resolution. This strategy was able to handle the difficult boundary conditions of galactic MOND Poisson solvers (see also \citealt{brada00a,brada00b,nipoti07a,llinares08,wu07}). We also introduced the ability to handle multiple populations with different particle masses as used in \cite{angus12} and also to update the positions and velocities, giving it the capability to handle evolving simulations. In the code there is a section that computes the QUMOND source density, right hand side of Eq~\ref{eqn:qumond2}, which we have repeated in the appendix. 

In that section, the gradients of the Newtonian potential are taken in the $x$, $y$ and $z$ Cartesian directions to find the Newtonian gravitational field at one-half cell from the node ($i$, $j$, $k$) in all six directions. On top of this, the magnitude of the $\nu$-function must be evaluated at each of these six locations. All we do is change $g_{z_1} \rightarrow g_{z_1}+g_{z,e}$ and $g_{z_2} \rightarrow g_{z_2}+g_{z,e}$ meaning the external field is exclusively in the $z$-direction. Here $g_{z,e}$ is the Newtonian external field. This has a knock-on effect for Eqs~\ref{eqn:qm2}-\ref{eqn:qm6}. One can see in Fig~\ref{fig:efe} a comparison of the $y$-direction (perpendicular to the external field) internal gravitational field of a satellite galaxy when the strength of the external gravitational field increases, up to the limiting point which is the Newtonian field. Clearly, the external field has a huge potential impact on the internal satellite dynamics.  


\subsection{Two component simulations setup}
\protect\label{sec:2comp}
\subsubsection{The code}
Recall that our simulations are performed using a Particle-Mesh grid that is centrally refining. The number of particles is fixed at $128^3$, which is sensible since there are roughly that many stars in the Carina dSph galaxy, given the luminosity is merely $4.4\times10^5\lsun$ (\citealt{mateo98}). Our coarsest grid has a box length of 4~Mpc in one dimension, with 65 cells per dimension. Our finest grid is 650~pc in length with still 65 cells. Thus, the spatial resolution improves greatly on the smaller grids.

In our case, we used a finest box of length 650~pc and a coarsest box of 4~Mpc. The smallest scale involved is roughly 10~pc - which is the size of the smallest cell on the finest grid, and of order the central mean particle/star separation. The largest scale involved is roughly 100~kpc, which is the current distance of Carina from the Milky Way. This means several of the coarsest grids are only used to accurately find the boundary conditions for the finer grids. Particles are advanced in their orbits depending on where they are in real space. If they are within 250~pc of the centre they are advanced using the gravity calculated on the finest grid. Between 250~pc and 500~pc they are advanced using the second finest grid and between 500~pc and 1~kpc we use the third finest grid, etc.

The particles are separated into 64 equal batches for input/output reasons. This made it convenient to assign 63-64ths of the particles to represent the mass distribution of the Carina dSph and the final 1/64th to represent the Milky Way. For the Milky Way, the spatial information is not carefully set, but rather is just spherically symmetric. The Milky Way's internal velocity information is not used because we use the dSph's frame of reference and so each time step, every particle representing the MW has its velocity and position updated by the same amount to mimic the true orbit the dSph would be executing around the MW. We use the dSph's frame of reference since we need the dSph to be centred on the most accurate part of the code. 
\subsubsection{The initial conditions}
\protect\label{sec:ics}
Since the gravity profile of a dSph embedded in an external field is non-axisymmetric, and its velocity distribution is anisotropic, it is not trivial to produce initial conditions that are stable. As an example, the procedure of \cite{brada00b} was to generate a spherical King profile from the distribution function for an isolated, isotropic dSph in deep MOND. They then increased the magnitude of the external field gradually until it attained the value the field should have at apocentre. This includes the external field in a shrewd way, but unfortunately the observations of \cite{walker07} and Fig~\ref{fig:m2l_pdf} have subsequently shown the dSphs require anisotropic velocity distributions. Also, King models produce poor fits to the surface brightness of \cite{munoz06}. We opted against following the procedure of \cite{brada00b} for a anisotropic distribution function because we have no guarantee that the anisotropy will be unaffected by the increasing external field.

We found that we could evolve to a relaxed dSph satisfying the surface brightness distribution of Fig~\ref{fig:munoz} starting from a spherical distribution with $\alpha=5.5$, $r_c=550~pc$, $\beta=-4$; the internal gravity was found using Eq~\ref{eqn:efe} and the radial velocity dispersion as a function of radius was found from solving the Jeans equation (see \S\ref{sec:jeansmond}). As well as the spatial distribution quickly changing as the simulation evolves, the velocity distribution changes too: from $\beta=-4$ to $\beta\sim-0.8$ in a timescale of 300~Myr. We used the standard rejection-sampling technique to define the positions of the particles within the dSph and assumed a Gaussian distribution for the velocities. We always initially offset the MW from the dSph along the $z$-axis, typically by +100~kpc, and give the velocity of the MW relative to the dSph along the +$y$-axis. The $x$-axis points out of the orbital plane.

We ran the simulations for up to $\sim$6~Gyr. We felt that simulating the dSphs for two full Galactocentric orbits would be sufficient to demonstrate the impact of the tides. We were also constrained by available computing resources.

\begin{figure*}
\centering
\subfigure{
\includegraphics[angle=0,width=8.0cm]{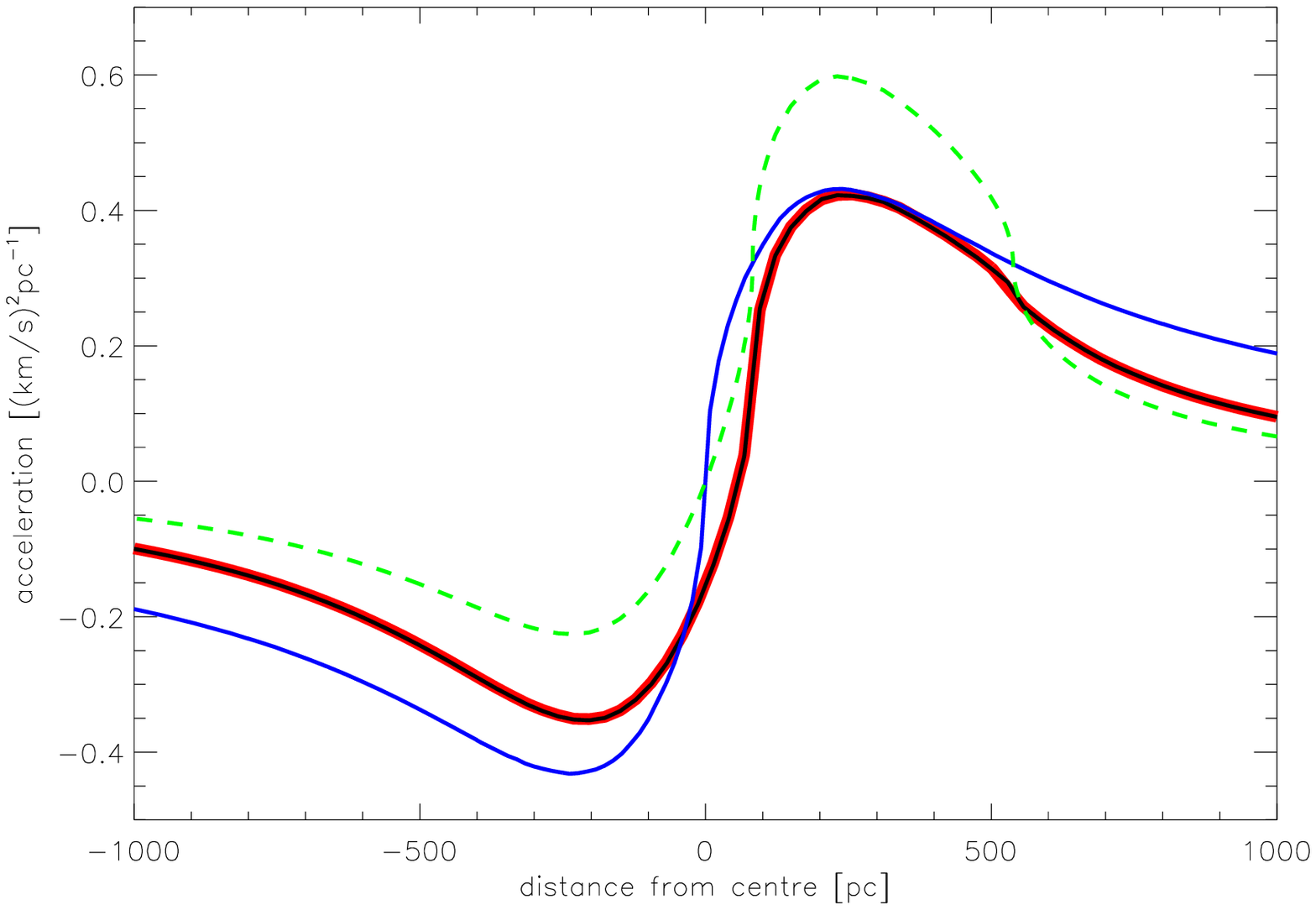}
}
\subfigure{
\includegraphics[angle=0,width=8.0cm]{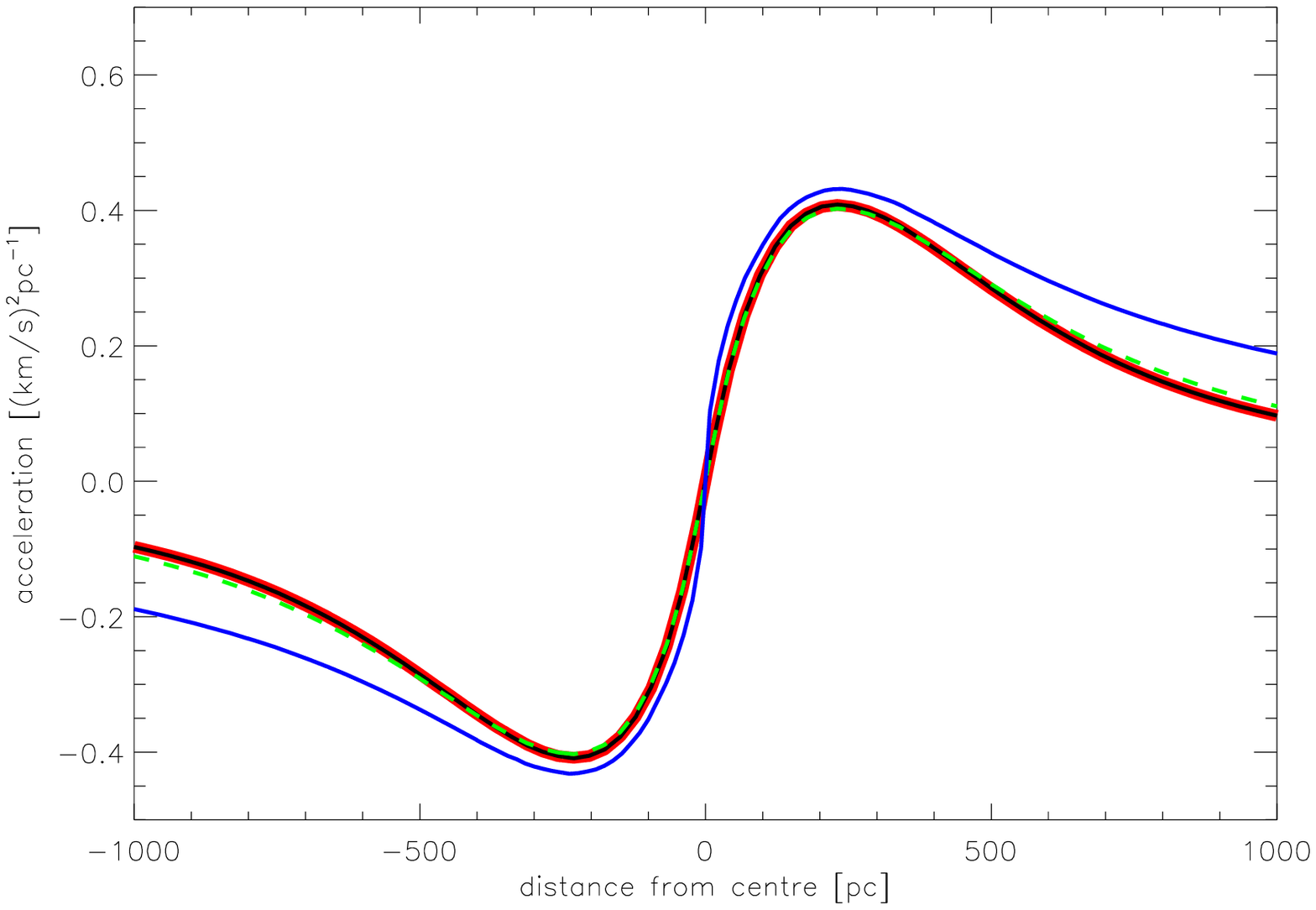}
}\\
\caption{Here we plot various computations of the internal gravity profile of Carina, using the parameters presented in \S\ref{sec:likejeans}. The left hand panel is in the $z$-direction (parallel to the external field) and the right hand panel is in the $y$-direction (perpendicular to the external field). In the left hand panel, the blue line shows the gravity profile for an isolated Carina with no external field. The positive distance from the centre is towards the Milky Way. The thick red curve is the $z$-direction gravity (towards or away from the Milky Way) using the two-component simulations with dSph and Milky Way. The black line, which lies on top of the thick red line, is for the dSph only simulation with the external field included via the boundary conditions. The dashed green line does not include the curl-field i.e. only solves Eq~\ref{eqn:efepar}.
In the right hand panel, the overlapping thick red and black lines are $y$-direction curves using the two-component simulation and one-component with boundary conditions respectively. The blue line is the gravity profile for an isolated Carina with no external field, which is larger in amplitude at all radii than the other curves. The dashed green line does not solve for the curl-field (i.e. only solves Eq~\ref{eqn:efe}).}
\label{fig:grav_prof}
\end{figure*}

\subsection{Comparison of external field parameterisations}
\protect\label{sec:efp}
There are clearly many parameterisations of the external field effect, as we noted at the beginning of this section. In order to demonstrate the differences in gravity profiles between them, we plot a series of curves together in Fig~\ref{fig:grav_prof} using the parameters presented in \S\ref{sec:likejeans}. The left hand panel is in the $z$-direction (parallel to the external field) and the right hand panel is in the $y$-direction (perpendicular to the external field). In the left hand panel, the blue line shows the gravity profile for an isolated Carina with no external field. The positive distance from the centre is towards the Milky Way. The thick red curve is the $z$-direction gravity (towards or away from the Milky Way) using the two-component simulations with dSph and Milky Way. The black line, which lies on top of the thick red line, is for the dSph only simulation with the external field included via the boundary conditions. The dashed green line does not include the curl-field i.e. only solves Eq~\ref{eqn:efepar}. This clearly is a very poor estimation of the gravity profile and should be avoided at all costs.

In the right hand panel, the overlapping thick red and black lines are $y$-direction curves (perpendicular to the external field) using the two-component simulation and one-component with boundary conditions respectively. The blue line is the gravity profile for an isolated Carina with no external field, which is larger in amplitude at all radii than the other curves. The dashed green line does not solve for the curl-field (i.e. only solves Eq~\ref{eqn:efe}). Inspecting the left hand panel's thick red (two-components) and black (boundary conditions) lines, the centre of gravity in the $z$-direction is not at Carina's mass weighted centre.

In summary, using the constant external field, boundary conditions gravitational field will be quite accurate as a probe of the instantaneous dynamics of a dSph, but using the curl-free solution will introduce large errors, especially in the direction of the external field. Regardless, the two-component simulations must be used to account for tidal fields which become more important the closer the dSph approaches the Milky Way.

\section{Comparison of simulated with observed velocity dispersions}
\protect\label{sec:compar}

\subsection{dSph stability}
\begin{figure}
\includegraphics[angle=0,width=8.0cm]{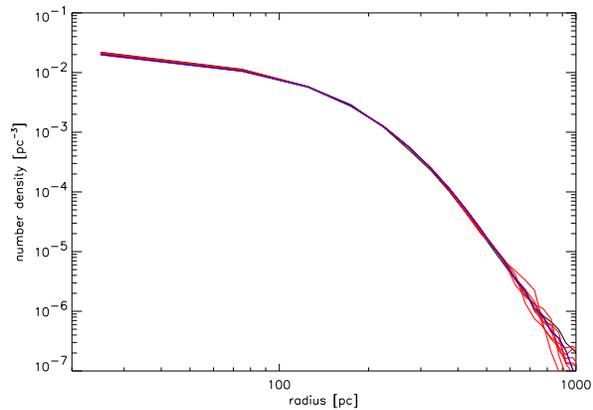}
\caption{Here we plot the evolution in the 3D density profile of an isolated Carina with no external field or simulated Milky Way over the period of 1~Gyr. The initial conditions are described in \S\ref{sec:ics}. The black line is the density after 300~Myr and the red lines show evenly spaced evaluations of the density over a 1~Gyr period. The blue line is the final density after 1~Gyr.}
\label{fig:dens}
\end{figure}

To check the stability of a dSph in isolation, we performed a simulation with no second component (no Milky Way nor external field) which lasted just over 1~Gyr with $M/L_V=5$. One Gyr is $\sim$50 dynamical times, and so any severe changes should already be prominent. In Fig~\ref{fig:dens} we plot the spherically averaged density profile of the dSph at regular intervals of time, up to just over 1~Gyr. Our first density curve (black line) is made after 300~Myr to ensure the dSph has had time to relax. The density changes only very slightly during this 1~Gyr period, suggesting the dSph is stable. The outer density fluctuates somewhat due to the slow ongoing leakage of particles. We found that time-steps of 0.01~Myr were required to reach convergence. Time-steps longer than this caused the dSph to dissolve on a 1~Gyr timescale. Later we show many plots of the surface density as a function of time for the dSph on orbits around the Milky Way.

\subsection{Sampling the mock catalogue}
Our comparison with the data of \cite{walker07} using our simulations is novel. In our re-binned projected velocity dispersion profile there are different numbers of stars per 50~pc bin. The number of stars per the central radius of the bin from $25~pc$ to $525~pc$ are 17, 43, 75, 79, 100, 83, 77, 67, 45, 33 and 29.

We create a mock catalogue of the line of sight velocities of a sample of stars in Carina by projecting
the 3D velocities of the particles within the simulated dSph along a direction between the simulated dSph and the solar position in the simulated MW.

Note that if this was for external dSphs, like one of the Andromeda dSphs (\citealt{mcconnachie12}), then this approach would not work since the satellite orbits are not Milky Way centric. It is essential to translate the positions and velocities because the velocity dispersion in the $z$-direction of the external field is lower than in the orthogonal directions. However, if we just use the untranslated $z$-direction velocity dispersions the velocity dispersions will not just change with Galactocentric radius, but also angle.

These above procedures were applied to all particles corresponding to the dSph. In the following, we sample a small number of the particles corresponding to the numbers of stars observed by \cite{walker07}.

Our next step is to randomly sample particles from the ensemble. With each sampled particle we calculate which projected radius bin it belongs to according to its newly translated projected radius, $R=\sqrt{x^2+y^2}$. Using this radius bin we add the square of the line of sight velocity, $v_z^2$, to the accumulated squared velocity dispersions in that bin. Our final condition is that the velocity of the particle with respect to the systemic velocity is less than $30~\kms$ since there is no star in the observed data with a relative velocity larger than this. We continue to sample a random sequence of particles until we have the same number of particles in each bin as we have stars noted at the start of this sub-section. We then divide the summed $z$-direction squared velocities by the number of stars in each particular bin and then take the square-root. Next, we compare this simulated projected velocity dispersion in each bin with the observed projected velocity dispersion in each bin and calculate the $\chi^2_{red}$. We repeat this process 10k times and each time we use a different random sequence of particles.

\subsection{Fraction of good fits}
\begin{figure*}
\centering
\subfigure{
\includegraphics[angle=0,width=8.0cm]{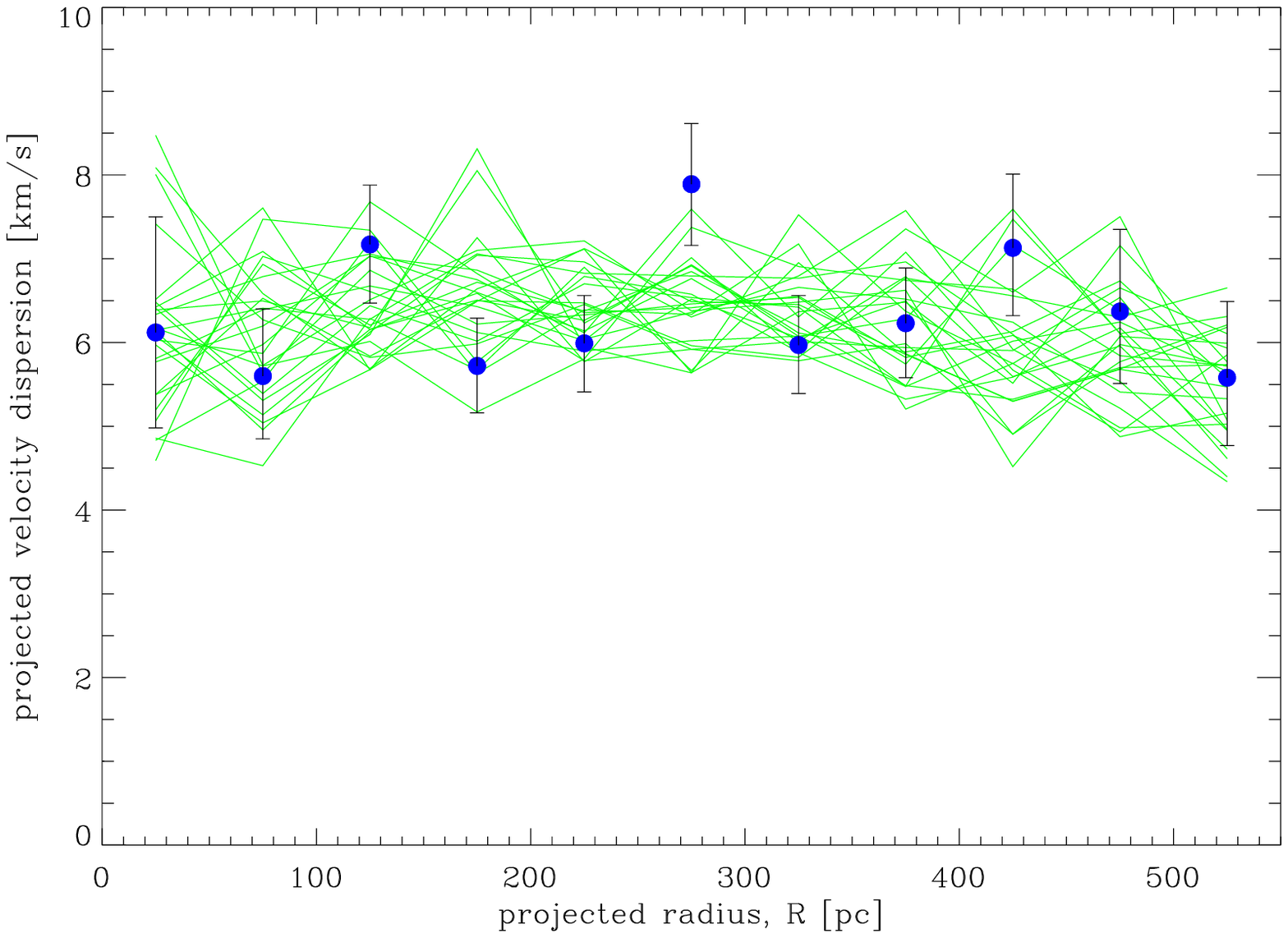}
}
\subfigure{
\includegraphics[angle=0,width=8.0cm]{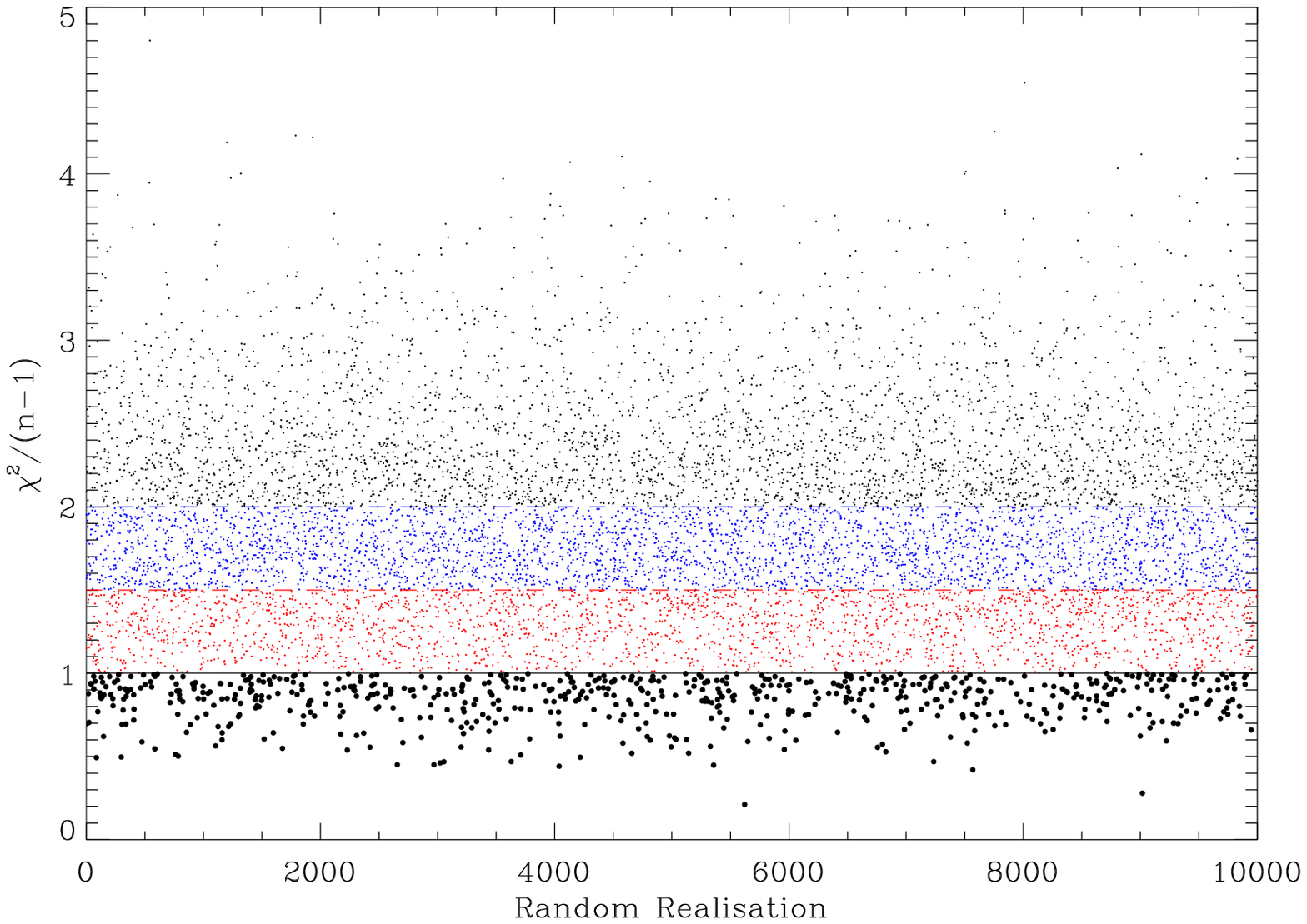}
}\\
\caption{In the left hand panel we plot our re-binning of the projected velocity dispersion data from Walker et al. (2007) with blue filled circles. Over-plotted with green lines are 25 random realisations of our  best fit static model of Carina as explained in \S\ref{sec:compar}. In the right hand panel we plot the $\chi^2_{red}$ distribution of 10k random realisations of the same model. $\chi^2_{red}$ levels are highlighted with different colours in order to clearly see the number of fits within the ranges $\chi^2_{red}<1$ (large black dots), $1<\chi^2_{red}<1.5$ (red dots), $1.5<\chi^2_{red}<2$ (blue dots) and $\chi^2_{red}>2$ (small black dots).}
\label{fig:chi_dist}
\end{figure*}

In the left hand panel of Fig~\ref{fig:chi_dist} we plot 25 (from the full 10k) random realisations of the simulated velocity dispersion profile of the static model (from \S\ref{sec:likejeans}, see also Fig~\ref{fig:munoz}), without allowing for any evolution. This is simply to show the variation in the simulated projected velocity dispersions even before tidal forces have influenced the dSph. One can see that at each radius bin there is a spread of velocity dispersions around the mean profile. Therefore, the possibility exists to have significantly larger or smaller velocity dispersions than the average in each bin. However, the presence of 11 radius bins precludes a velocity dispersion profile with a low amplitude from randomly producing a good fit.

In the right hand panel we plot $\chi^2_{red}$ for 10k random realisations of the same static model. $\chi^2_{red}=0.9$ is the best fit smooth curve to the re-binned data of \cite{walker07} (either with MOND or Newtonian gravity and a DM halo). There are, however, a number of realisations for which $\chi^2_{red}<0.5$, demonstrating the potential to have a significantly better match to the data than the smooth fit allows.

The fraction of realisations with $\chi^2_{red}<1$ is 0.06. For the rest of the paper we use this as our figure of merit because it gives a robust likelihood of Carina at its given orbital position having a good match to the data. $\chi^2_{red}<1$ was simply a suitable number that would yield a statistically useful return after 10k realisations. Using a lower $\chi^2_{red}$ threshold would lead to low number statistics and a much higher one, say $\chi^2_{red}<1.5$ or 2 (the dashed red and blue lines respectively in Fig~\ref{fig:chi_dist}), would contain information about poorer fits that we are less interested in. It remains that a realisation which generally produces a higher number of $\chi^2_{red}<0.5$ than another one, will produce a higher fraction of $\chi^2_{red}<1$ than that other realisation. We are aware that the correlation of the errors in each radius bin means we have over-estimated those errors, however, since we only use the $\chi^2_{red}$ to compare fits in a relative sense, we do not see this as a problem.

\section{Results}
Our goal is to determine whether tidal effects from the Milky Way have a meaningful influence on the Carina dSph and if they are conducive to lowering the inferred mass-to-light ratio relative to that garnered from Jeans modelling. It's also crucial to confirm that dSphs like Carina are stable for several orbits with realistic orbital parameters. To accomplish this, we ran a series of simulations of a Carina-like dSph orbiting the Milky Way with initial conditions as described in \S\ref{sec:2comp}. We use a variety of total stellar masses for the dSph and orbital parameters with respect to the Milky Way.

The mass-to-light ratio that this stellar mass corresponds to changes with time since the dSph begins to lose a small fraction of particle mass due to tidal stripping as soon as the simulations start. This is because the initial conditions quickly transform to an equilibrium distribution to which not all particles are bound. The mass of our particles of course do not change, but we renormalize their corresponding luminosity while the dSph loses mass in order to fit its observed total luminosity.

The orbital parameters that we discuss refer to the initial radial distance from the Milky Way and the initial velocity: both radially from and tangentially to the Milky Way. Obviously, the closer the pericentric distance from the Milky Way, the more mass will be stripped and the higher the mass the dSph has, the more resilient it will be to tides.

Since Carina currently appears to be approaching its apocentre, with current distance of $\sim100~kpc$ (see \citealt{piatek03,metz08}), this is where we must always make our comparison with the other simulations - even if in the simulation the apocentre is larger than $100~kpc$. The reason for this is that increasing the radial distance from the Milky Way decreases the external gravitational field acting on Carina and thus increases the boost to the internal gravity due to MOND. This would allow smaller mass-to-light ratios to be consistent with the observed dynamics than possible at $100~kpc$.

\subsection{Measured proper motions}

The Carina dSph has measured proper motions from the observations and analysis of \cite{piatek03}. They use the proper motions, estimated from two separate stellar fields within Carina, along with the well measured line of sight velocity to estimate the radial ($V_r$) and tangential ($V_t$), with respect to the Milky Way centre, orbital velocities. The first field gives $V_r=18\pm32~\kms$ and the second field gives $V_r=22\pm36~\kms$ and are therefore consistent with each other and produce a weighted mean $V_r=20\pm24~\kms$. The tangential velocity for the first field is $V_t=40\pm53~\kms$ and for the second field is $V_t=140\pm59~\kms$, which is a considerable difference that barely allows an overlap within the errors. The weighted mean tangential velocity is therefore $V_t=85\pm39~\kms$, but not much credence should be given to the formal error since only two, vastly differing, fields have been measured.

Following up on this measurement, \cite{metz08} corrected for the advanced charge transfer inefficiencies of the Space Telescope Imaging Spectrograph (\citealt{bristow05}) and found the updated weighted means to be $V_r=22\pm3~\kms$ and $V_t=120\pm50~\kms$.

We do not consider the radial velocity as being important in our simulations. Our reasoning is that it is small relative to the tangential velocity and the radial velocity only influences the apocentre, whereas the tangential velocity sets the pericentre. Given that only the pericentre sets the impact from tides, we can safely ignore the small radial velocity.
\subsection{Fraction of good fits as function of orbit}

\begin{figure*}
\includegraphics[angle=0,width=17.0cm]{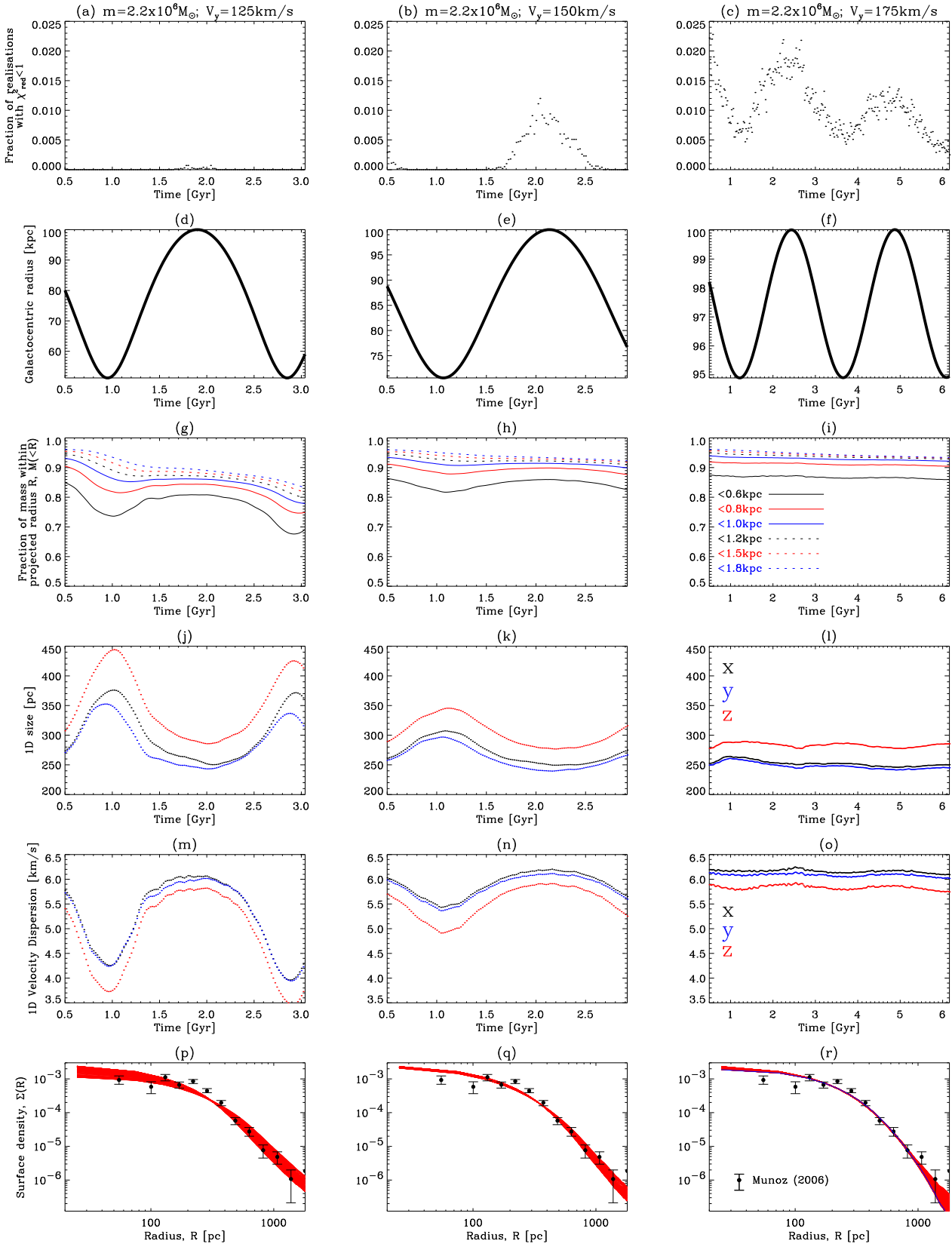}
\caption{\tiny Here we plot six different quantities (each row) for a different orbit (each column) using the same initial conditions and the same initial dSph mass (here $m=$2.2$\times10^6\msun$). The three different orbits are defined by their initial tangential velocity ($V_y$) which is in the direction perpendicular to the initial separation between the dSph and the Milky Way of $100~kpc$: from left to right $V_y=125$, $150$ and $175~\kms$. (ii) The Galactocentric distance as a function of time is the second row of plots and (i) the top row shows the fraction of random realisations of the projected velocity dispersion which, when compared with the observed one, yield $\chi^2_{red}<1$ as a function of time. (iii) The third row gives the fraction of projected mass within radial shells of 0.6, 0.8, 1.0, 1.2, 1.5 and 1.8~kpc as a function of time. (iv) The fourth row give the 1D RMS size in each of the three directions ($x$, $y$ and $z$; where $z$ is the line of sight) and (v) gives the 1D RMS velocity. (vi) The final row shows surface density profiles (red lines) for evenly spaced snapshots in time. The normalisation is the same for every snapshot. The blue line in panel (r) is the initial surface density at the start of the simulation. Also over-plotted (filled circles with error bars) is the observed surface density of stars from Munoz et al. (2006).}
\label{fig:ml5}
\end{figure*}

\begin{figure*}
\includegraphics[angle=0,width=17.0cm]{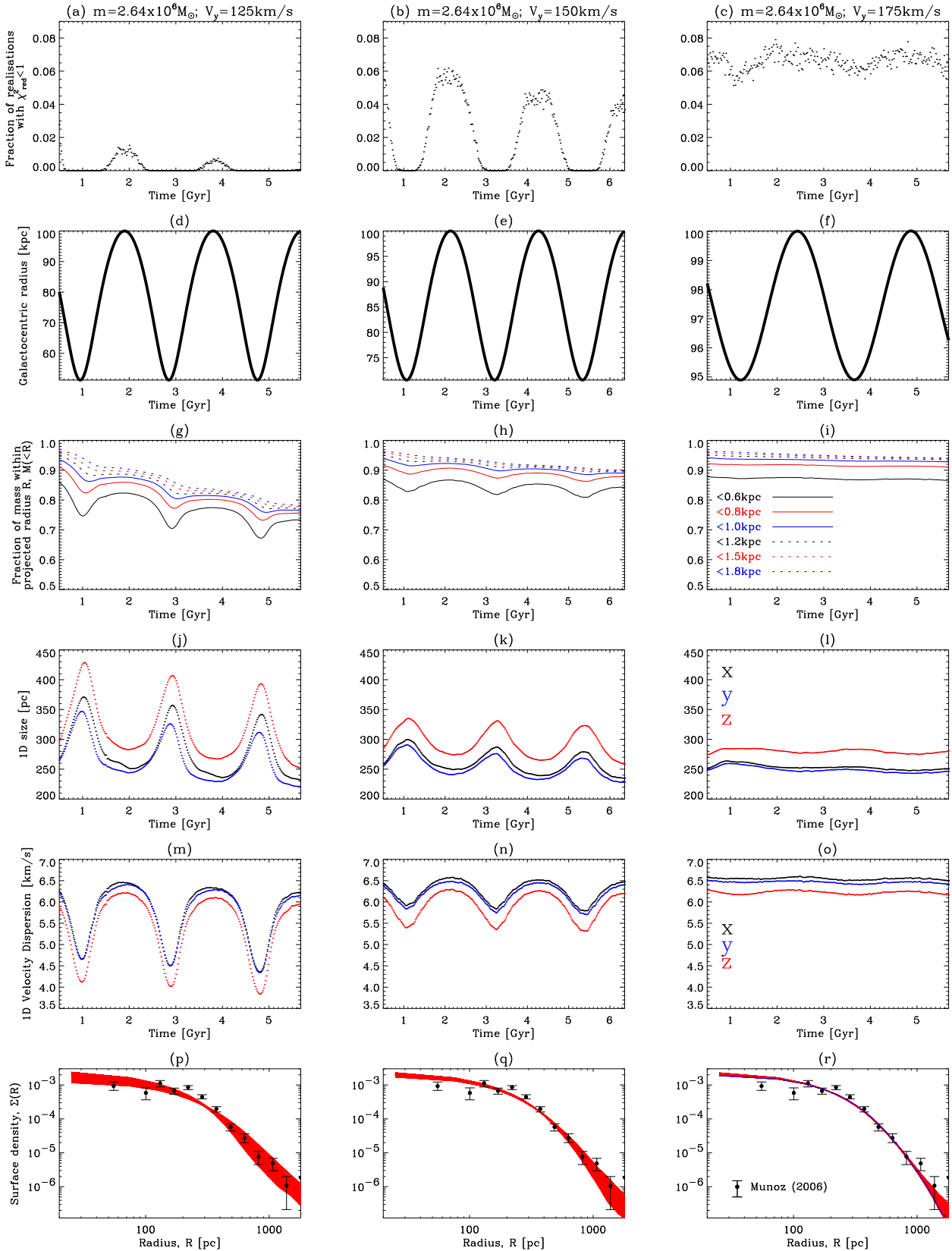}
\caption{As per the caption of Fig~\ref{fig:ml5}, but for $m=$2.64$\times10^6\msun$}
\label{fig:ml6}
\end{figure*}

\begin{figure*}
\includegraphics[angle=0,width=17.0cm]{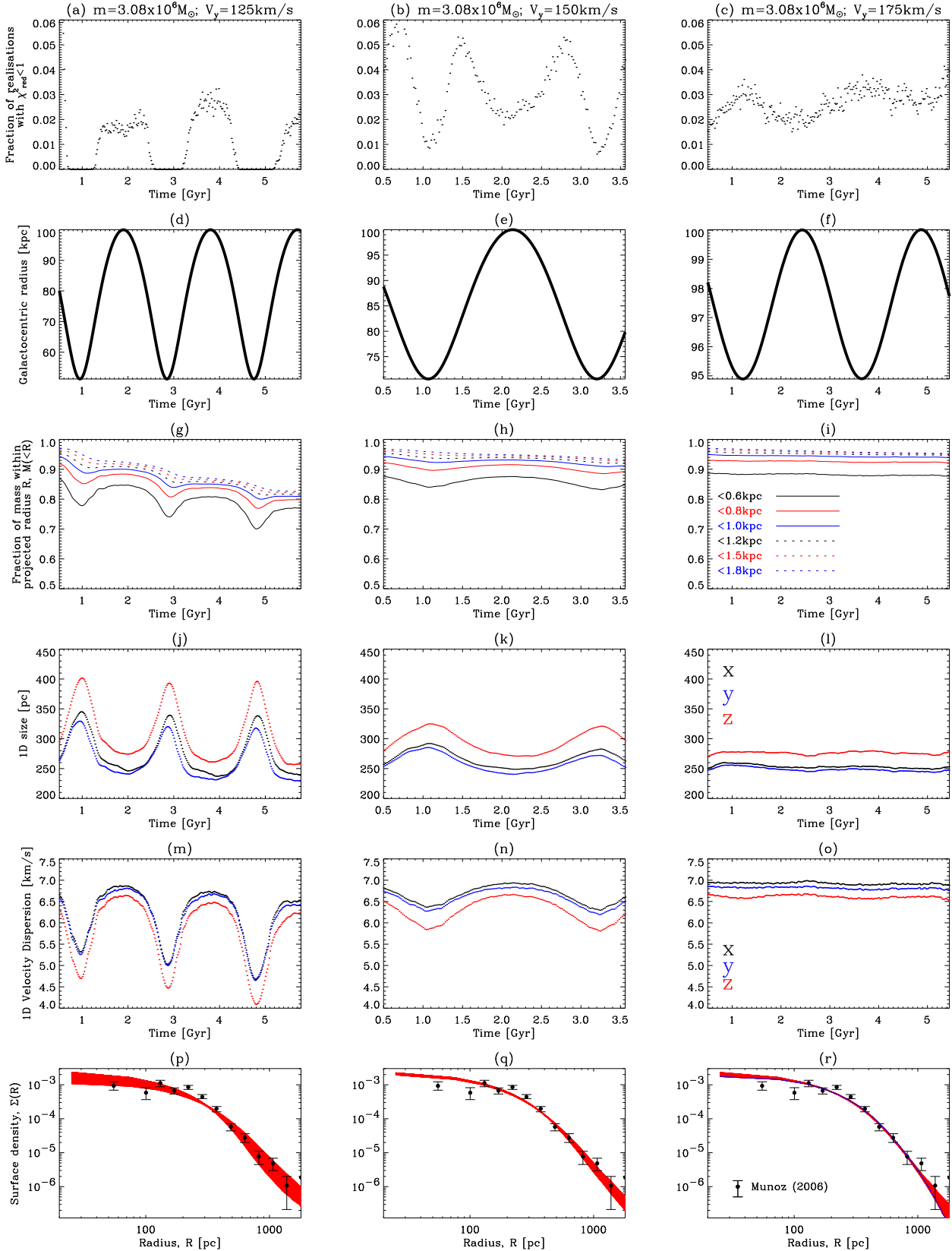}
\caption{As per the caption of Fig~\ref{fig:ml5}, but for $m=$3.08$\times10^6\msun$}
\label{fig:ml7}
\end{figure*}

To investigate the suitability of a given Carina stellar mass and orbital path to matching the projected velocity dispersion data of \cite{walker07}, we ran three simulations with different orbital parameters for each of the total Carina stellar masses $m=$2.2, 2.64 and 3.08$\times10^6\msun$, which correspond to 5, 6 and 7 times the luminosity with mass-to-light ratio of unity. The three different orbital parameters are simply the initial tangential velocity relative to the initial offset along the $z$ axis of $100~kpc$. These were $V_y=125$, 150 and 175~$\kms$. $V_y=175~\kms$ leads to an almost circular orbit with pericentre of $\sim$95~kpc and orbital period of $\approx$2.4~Gyr. $V_y=125$ and 150~$\kms$ give pericentres of 51 and 71~$kpc$, respectively and orbital periods of $\approx$1.9 and 2.1~Gyr. In Fig~\ref{fig:ml5} one can see 6 rows of plots for the $m=$2.2$\times10^6\msun$ simulations, where the left, middle and right hand columns refers to the initial tangential velocities $V_y=125$, 150 and 175~$\kms$ respectively. The Figs~\ref{fig:ml6} and ~\ref{fig:ml7} show the same plots but for the $m=$2.64 and 3.08$\times10^6\msun$ simulations.

\begin{figure*}
\includegraphics[angle=0,width=17.0cm]{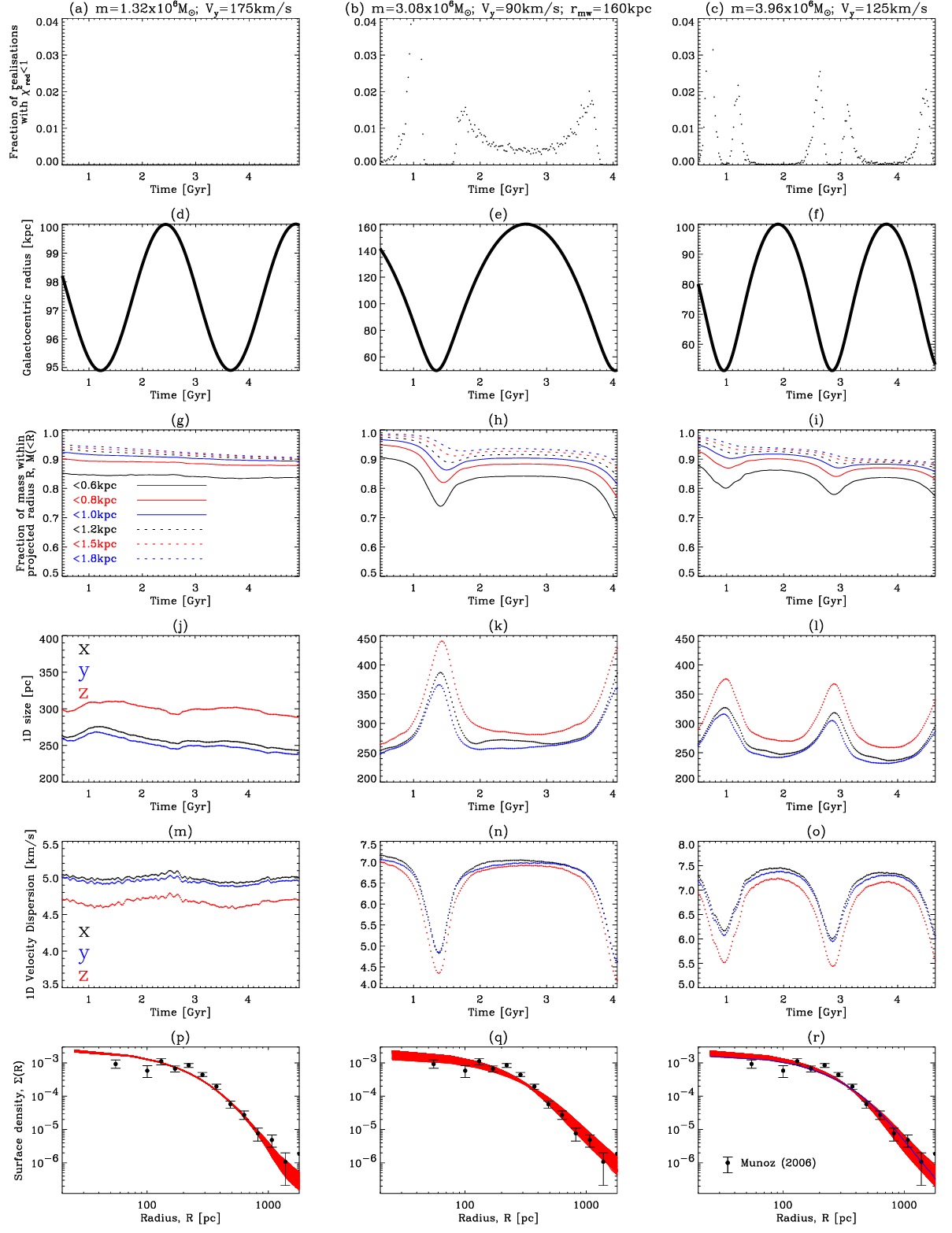}
\caption{As per the caption of Fig~\ref{fig:ml5}, but for three odd models. From left to right $m=$1.32$\times10^6\msun$, $V_y=175~\kms$; $m=$3.08$\times10^6\msun$, $V_y=90~\kms$, $r_{mw}=160~kpc$; $m=$3.96$\times10^6\msun$, $V_y=125~\kms$.}
\label{fig:mlodd}
\end{figure*}

For Figs~\ref{fig:ml5}-\ref{fig:mlodd}, the top row (panels a-c) shows, as a function of time, the fraction of random realisations of the projected velocity dispersion which, when compared with the observed one, yield $\chi^2_{red}<1$. This is the most important plot, which shows whether this particular combination of total stellar mass and tangential velocity will well reproduce the observed projected velocity dispersions (this procedure is discussed in more detail in \S\ref{sec:compar} and Fig~\ref{fig:chi_dist}). The second row (panels d-f) in Fig~\ref{fig:ml5} gives Galactocentric distance as a function of time for the simulations.

Looking specifically at Fig~\ref{fig:ml6} ($m=$2.64$\times10^6\msun$) for the $V_y=175~\kms$ (right hand column) simulations, one can see from panel (c) there are not dramatic changes in the fraction of good fits with time because the orbital distance from the MW does not change significantly during the orbit and we know how crucial the Galactocentric radius is to the internal dynamics in MOND (e.g. Fig~\ref{fig:efe}). On the other hand, the simulations with $V_y=125$ and 150~$\kms$  (left and middle columns of Fig~\ref{fig:ml6} respectively) reach significantly smaller pericentres and this means they are exposed to stronger external gravities which cause the internal gravities to be reduced. In panel (c) of Figs~\ref{fig:ml5} and \ref{fig:ml7} (i.e. for  $V_y=175~\kms$), the fraction of good fits appears to vary more with the orbit than for  Fig~\ref{fig:ml6} panel (c). Actually, the magnitude of the change is similar for all three, it is simply that for Fig~\ref{fig:ml6} ($m=$2.64$\times10^6\msun$) the variation is relative to a larger number.

Another important thing is that on the smaller pericentre orbits, more mass is stripped and this is shown in the third row of plots (Fig~\ref{fig:ml5}-\ref{fig:mlodd} panels g-i) where the projected enclosed mass in several shells is given. The outermost shell is the projected mass within R=1.8~kpc and should contain most of the bound mass. For the $V_y=125~\kms$ simulation of Fig~\ref{fig:ml6} (panel g; $m=$2.64$\times10^6\msun$), the mass in the outer shell (R$<$1.8~kpc) drops from more than 95\% to less than 80\% after three orbits, but the $V_y=175~\kms$ (panel i) simulation only loses about 2\% of the mass in that shell.

The reduction of the internal gravity, due to the varying external field strength, can be clearly seen in the fourth and fifth rows of plots of Fig~\ref{fig:ml6} (panels j-o) which show the 1D RMS sizes in the three orthogonal directions and the 1D RMS velocities. For $V_y=175~\kms$, the RMS velocity (panel o) in each of the three directions is very constant as are the RMS sizes (panel l). However, for $V_y=125$ and 150~$\kms$, the sizes (panels j and k) and RMS velocities (panels m and n) change according to the orbit. This is why the fraction of good fits is also a function of time.

The final row of plots for Figs~\ref{fig:ml5}-\ref{fig:mlodd} (panels p-r) show the surface density profiles for evenly spaced snapshots in time. The normalisation is the same for every snapshot. The blue line in panel (r) is the initial surface density at the start of the simulation. There is very little change in shape for any of the orbits, but the $V_y=175~\kms$ (panel r) simulation is particularly constant and this demonstrates the stability of the dSphs in MOND and our simulations. We show for comparison the surface density of stars found by \cite{munoz06}. One final point to take from these bottom three rows of plots in Figs~\ref{fig:ml5}-\ref{fig:mlodd} is how the $z$-direction size (panels j-l) and RMS velocity (panels m-o) are different to the similar $x$ and $y$ direction RMS velocities and sizes. This is obviously due to the direction of the external field which points along the $z$ direction at every time step because in the post simulation analysis we rotate our frame of reference such that the $z$ direction always points towards the Milky Way. This stretching along the external field direction is a well known effect is investigated by \cite{milgrom86,zhaot06,wu08} and will be the topic of future study.

The top rows of Figs~\ref{fig:ml5}-\ref{fig:mlodd} (panels a-c) show the suitability of the combination of mass and tangential velocity $V_y$. Here it is important that we only compare the fractions of good fits at apocentre (100 ~kpc) and clearly the highest fraction of good fits is found using a mass of $m=$2.64$\times10^6\msun$ and $V_y=175~\kms$, which easily gives 0.07 - slightly larger than found from the sampling of the isolated model (0.06). Using $m=$3.08$\times10^6\msun$ and $V_y=175~\kms$ gives roughly 0.03 and $m=$2.2$\times10^6\msun$ and $V_y=175~\kms$ is roughly 0.02 after one orbit and much lower at 0.012 after two orbits. The reason $m=$3.08$\times10^6\msun$ does not work as well as $m=$2.64$\times10^6\msun$ is because it is too massive and leads to excessively high velocity dispersions $\sim 6.6~\kms$ in the $z$-direction (Fig~\ref{fig:ml7}, panel o). This can clearly be seen because the agreement improves at 94~kpc over 100~kpc where the larger external field reduces the velocity dispersion. For $m=$2.2$\times10^6\msun$, the dSph is not massive enough and generates too small velocity dispersions $\sim 5.8~\kms$ in the $z$-direction  (Fig~\ref{fig:ml5}, panel o). Furthermore, even on the nearly circular orbit with $V_y=175~\kms$, it is stripped gradually by tidal forces and this means that the number of good fits (with $\chi^2_{red}<1$) to the observed velocity dispersions decreases during each orbit.

\begin{figure}
\includegraphics[angle=0,width=8.0cm]{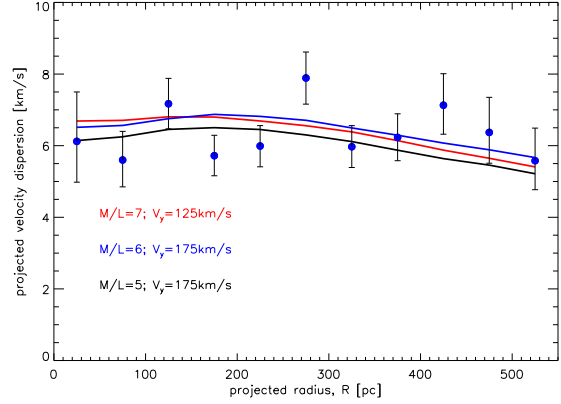}
\caption{Here we plot the projected velocity dispersions using all particles of three simulations ($m=$2.2$\times10^6\msun$ and $V_y=175~\kms$; black), ($m=$2.64$\times10^6\msun$ and $V_y=175~\kms$; blue) and ($m=$3.08$\times10^6\msun$ and $V_y=125~\kms$; red) after two, two and three full orbits respectively. The data points are our re-binned projected velocity dispersions using the data of Walker et al. (2007).}
\label{fig:anis}
\end{figure}

Using $m=$2.64$\times10^6\msun$ and $V_y=150~\kms$ (Fig~\ref{fig:ml6} panel b) has a high fraction of good fits after one orbit ($\sim$0.06), but this drops after each orbit because mass is stripped (panel h) reducing the gravitational field and velocity dispersions. With $m=$2.64$\times10^6\msun$ and $V_y=125~\kms$ (Fig~\ref{fig:ml6} panel g), too much mass is stripped after one orbit and a very low number of good fits is left (panel a). The lower $V_y$s are also ruled out for $m=$2.2$\times10^6\msun$. For $m=$3.08$\times10^6\msun$ and $V_y=125~\kms$ (Fig~\ref{fig:ml7}), the correct amount of mass is stripped (panel g) by the end of the second orbit, but the fraction of good fits (panel a) is still quite mediocre ($\sim$0.02). This appears to be a result of the tidal forces changing the velocity anisotropy. To clarify this, in Fig~\ref{fig:anis} we plot the projected velocity dispersion using all particles (not just the number of observed stars) for three simulations after roughly 5~Gyr each. The first curve (red) is for the simulations with $m=$3.08$\times10^6\msun$ and $V_y=125~\kms$ and is taken after three full orbits. The other two curves both use $V_y=175~\kms$ after two full orbits, but have different masses: $m=$2.2 and 2.64$\times10^6\msun$ (black and blue curves respectively). Clearly, the $m=$2.2 and 2.64$\times10^6\msun$ curves have the same shape, but the $m=$2.64$\times10^6\msun$ curve has a larger amplitude and significantly better $\chi^2_{red}$, however, the $m=$3.08$\times10^6\msun$ curve has a different shape because it has a more isotropic velocity anisotropy. Therefore, the plunging orbit exposes the $m=$3.08$\times10^6\msun$ dSph to tides that distort the velocity distribution towards less tangentially biased orbits which makes it slightly less consistent with the observed projected velocity distributions, according to the fraction of good fits and half as likely to produce the observed velocity dispersions.

Finally, we have three extra simulations for which we plot (in Fig~\ref{fig:mlodd}) the relevant quantities, as per Figs~\ref{fig:ml5}-\ref{fig:ml7}. The $m=$1.32$\times10^6\msun$ plot (Fig~\ref{fig:mlodd} panel a) shows that if $m=$1.32$\times10^6\msun$ there is no likelihood of being consistent with the observed projected velocity dispersions. The right hand column is for $m=$3.96$\times10^6\msun$ with $V_y=125~\kms$ and also shows (panel a) that at apocentre (100~kpc), there is no likelihood of it being consistent. The central column is for a simulation with $m=$3.08$\times10^6\msun$ that does not start at the fiducial 100~kpc, but rather at 160~kpc and plunges to $\sim 50~kpc$. Both on the inbound and outbound sections of the orbit at 100~kpc, the fraction of good fits (panel b) is mediocre at $\sim 0.015$. This is simply because the mass is too high. We do not have a similar simulation for this orbit and $m=$2.64$\times10^6\msun$.

Therefore, the most appropriate model appears to be $m=$2.64$\times10^6\msun$ and the closer the orbit is to circular the better the match. Conversely, the more plunging the orbit, the more the velocity anisotropy is transformed towards less tangentially biased orbits creating a poorer match. An important point to bear in mind is that this initial mass $m=$2.64$\times10^6\msun$ actually means a final mass which is slightly lower due to the stripped stars/particles. The fraction of stars left bound to the dSph depends on the size of the shell we consider, but is somewhere between 90 and 95\%. This means a final $M/L$ of between 5.4 and 5.7, which is close to the best fit value found using Jeans analysis (Fig~\ref{fig:m2l_pdf}). On top of this is the uncertainty in the observed luminosity of Carina. The reason the preferred $M/L$ is slightly larger than the maximum likelihood of 5.2 from Jeans modelling is that the external field causes elongation in the $z$-direction and likewise causes the velocity dispersion to be smaller than in the two orthogonal directions. This effect is not taken into account in the Jeans modelling. 

\begin{figure}
\includegraphics[angle=0,width=8.0cm]{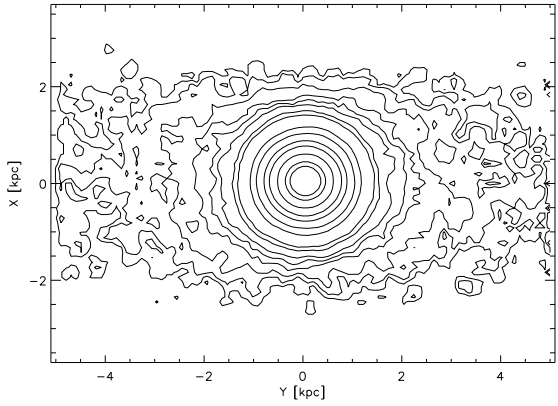}
\caption{Here we plot projected density contours for a simulation with $m=$2.64$\times10^6\msun$ and $V_y=175~\kms$ after 5~Gyr. Up and down is the $x$-axis (out of the orbital plane), left and right is the $y$-axis and the $z$-axis is along the line of sight.}
\label{fig:batt}
\end{figure}

\cite{battaglia12} have demonstrated that the projected stellar distribution of Carina has tidal tails which suggest ongoing harassment of the dSph from the Milky Way. In Fig~\ref{fig:batt} we show contours of the projected particle distribution for Carina on a near circular orbit after 5~Gyr. The tidal tails outside the circular isodensity contours of the bound particles lie in the plane of the orbit and occur naturally in MOND even on a near circular orbit at 100~kpc. Whether the same is true for such a distant orbit in CDM simulations remains to be seen.

\section{Conclusion}
Here we have run a suite of MOND N-body simulations of a dSph like Carina with various total masses ($m=$1.32, 2.2, 2.64, 3.08 and 3.96$\times10^6\msun$) and orbital paths around the Milky Way. We have shown that they are stable and long lived on nearly circular orbits at 100~kpc regardless of mass ($\ge m=$1.32$\times10^6\msun$) and even on orbits that plunge to 50~kpc. However, the model most likely to give a good fit to the observed projected velocity dispersions is one with an initial $m=$2.64$\times10^6\msun$, which means a $M/L$ in the range of 5.4 and 5.7 after two orbits ($\sim 5 Gyr$). The more circular the orbit, the less disturbed the internal velocity distribution is. This is important because the observations require substantially negative (tangentially biased) velocity anisotropies. After plunging orbits, the velocity anisotropy becomes slightly more radially biased, reducing agreement with the observations. Considering that a $M/L$ in the range of 5.4 and 5.7 is potentially at odds with stellar populations synthesis models, we considered a model with $m=$2.2$\times10^6\msun$, which after a single orbit corresponds to a $M/L$ between 4.5 and 4.7. This model has a likelihood of matching the observations that is roughly 3.5 times smaller than the model with $M/L$ between 5.4 and 5.7. This range of mass-to-light ratios is slightly above those found from basic Jeans analysis because the isopotential contours are stretched (see e.g. \citealt{milgrom86,zhaot06,wu08}) in the direction away from the Milky Way (which coincides here with our line of sight) due to the external field effect. This leads to a stretching of the dSph along the line of sight, relative to the plane perpendicular, and a reduction of the velocity dispersions.

As for the compatibility of different orbits, it would appear that after two orbits with initial $V_y=125~\kms$, the lower masses $m=$2.2 and 2.64$\times10^6\msun$ are not capable of generating a sizable fraction of good fits. $m=$2.2$\times10^6\msun$ would give less than 0.001, $m=$2.64$\times10^6\msun$ less than 0.01, but $m=$3.08$\times10^6\msun$ would produce roughly 0.03. This is because mass has been stripped leaving the true $M/L$ after two orbits to be somewhere between 5.9 and 6.2. Using $m=$2.64$\times10^6\msun$ after only one orbit with $V_y=125~\kms$ gives a fraction of good fits of only 0.015 with a true $M/L$ between 5.1 and 5.4. So the best fit $M/L$ for $V_y=125~\kms$ is likely somewhere between these two limits. However, it will probably still be somewhat less likely than the more circular orbits since the tides adversely affect the velocity anisotropy. For the intermediate orbit with $V_y=150~\kms$, $m=$2.64$\times10^6\msun$ leads to a fraction of 0.04 good fits after three full orbits with a true $M/L$ of $\sim$5.3-5.4. Therefore, for $V_y \ge 125~\kms$ the preferred $M/L$ remains fairly constant (5.3-5.7), but obviously on the more plunging orbits mass is more rapidly stripped and thus it is required that the current $M/L$ is in this range, not the initial one. A parallel observation is that the fraction of stripped mass during a period of almost half the age of the Universe is not more than half on any of the simulated orbits. Therefore, it must be the case that the dSph was formed with a mass very close to its current one and this is likely also true in the CDM paradigm.

Although the preferred $M/L$ is between 5.3 and 5.7, there is still a reasonable probability that the $M/L$ is lower than 5. From the various orbits it would seem that even on a near circular orbit, panel (c) of Fig~\ref{fig:ml5} shows (after one orbit) that $M/L\sim4.8$ is more than three times less likely than the best model. Panel (a) of Fig~\ref{fig:ml5} suggests that on an orbit with a 50~kpc pericentre, a $M/L\sim4.5$ has an insignificant probability of producing a good fit.

A larger sample of stellar line of sight velocities might subdue the errors here to distinguish between different mass-to-light ratios. Therefore, higher precision proper motions, larger samples of stars, ultra-precise photometry for the total luminosity, and more sophisticated and reliable stellar population synthesis models, as well as a full-fledged treatment of binaries for dwarf spheroidals would be enormously useful for future studies. Another factor that should be built in to future studies of Carina, is the possibility for triaxiality in the 3D stellar distribution. This must be an important factor because all dSph surface brightnesses are observed to be moderately elliptical (\citealt{irwinhatz}).

\section{acknowledgements} The authors are indebted to the referee for considerably improving the content and readability of the paper. GWA is a postdoctoral fellow of the FWO Vlaanderen (Belgium). Part of the research was carried out while GWA was a postdoctoral fellow supported by the Claude Leon Foundation. AD acknowledges partial support from the INFN grant Indark (PD51) and from the grant \verb'Progetti di Ateneo/CSP TO_Call2_2012_0011' 'Marco Polo' of the Universita' di Torino.

\section{appendix: Particle-Mesh external field inclusion}

Assume we want the QUMOND source (right hand side of Eq~\ref{eqn:qumond2}) at cell ($i$,$j$,$k$) of a Cartesian grid ($x$,$y$,$z$), then we need to define the gravity at various points surrounding it. If we use unit length grid cells then
\bey
\protect\label{eqn:qm1}
\nonumber g_{x_2} = \phi_{i+1,j,k} - \phi_{i,j,k}\\
\nonumber g_{x_1} = \phi_{i,j,k} - \phi_{i-1,j,k}\\
\nonumber g_{y_2} = \phi_{i,j+1,k} - \phi_{i,j,k}\\
\nonumber g_{y_1} = \phi_{i,j,k} - \phi_{i,j-1,k}\\
\nonumber g_{z_2} = \phi_{i,j,k+1} - \phi_{i,j,k}\\
g_{z_1} = \phi_{i,j,k} - \phi_{i,j,k-1}
\eey
These are the values of the Newtonian gravitational field at half a cell from ($i$,$j$,$k$) in the three orthogonal directions and $\phi$ is the Newtonian gravitational potential. Similarly, for these six points we must find the value of the $\nu$ function. Surrounding the point $x_2$ we use the dummy variable $\omega$ which is just the gravitational field in each of the orthogonal directions at a half cell from ($i$,$j$,$k$)
\bey
\protect\label{eqn:qm2}
 \omega_{x_2}&=&\phi_{i+1,j,k} - \phi_{i,j,k}\\
\nonumber 4\omega_{y_2}&=&\phi_{i+1,j+1,k} + \phi_{i,j+1,k} - \left(\phi_{i+1,j-1,k} + \phi_{i,j-1,k} \right)\\
\nonumber 4\omega_{z_2}&=&\phi_{i+1,j,k+1} + \phi_{i,j,k+1} - \left(\phi_{i+1,j,k-1} + \phi_{i,j,k-1} \right)
\nonumber
\eey
and surrounding $x_1$
\bey
\protect\label{eqn:qm3}
 \omega_{x_1}&=&\phi_{i,j,k} - \phi_{i-1,j,k}\\
\nonumber 4\omega_{y_1}&=&\phi_{i,j+1,k} + \phi_{i-1,j+1,k} - \left(\phi_{i,j-1,k} + \phi_{i-1,j-1,k} \right)\\
\nonumber 4\omega_{z_1}&=&\phi_{i,j,k+1} + \phi_{i-1,j,k+1} - \left(\phi_{i,j,k-1} + \phi_{i-1,j,k-1} \right)
\nonumber
\eey
which gives
\bey
\protect\label{eqn:qm4}
\nonumber \kappa_{x_2}=(a_o)^{-1}\sqrt{\omega_{x_2}^2+\omega_{y_2}^2+\omega_{z_2}^2}\\
 \kappa_{x_1}=(a_o)^{-1}\sqrt{\omega_{x_1}^2+\omega_{y_1}^2+\omega_{z_1}^2}.
\eey
which are the arguments for the $\nu$ function at each half cell from ($i$,$j$,$k$) in the three orthogonal directions. These are accompanied by $\kappa_{y_2}$, $\kappa_{y_1}$, $\kappa_{z_2}$ and $\kappa_{z_1}$ found in a similar way. From this we must find 
\bey
\protect\label{eqn:qm5}
\nonumber \nu_{x_2}=\nu(\kappa_{x_2})\\
\nu_{x_1}=\nu(\kappa_{x_1})
\eey
and $\nu_{y_2}$, $\nu_{y_1}$, $\nu_{z_2}$, $\nu_{z_1}$. This finally leaves us with the QUMOND source density in cell ($i$,$j$,$k$) given by 

\beq
\protect\label{eqn:qm6}
\grad \cdot \left[ \nu(y) \grad\Phi_N \right]=\nu_{x_2}g_{x_2} - \nu_{x_1}g_{x_1} + \nu_{y_2}g_{y_2} - \nu_{y_1}g_{y_1}+ \nu_{z_2}g_{z_2}- \nu_{z_1}g_{z_1}.
\eeq
A good visualisation of the geometry can be found in \cite{tcevol}, \cite{llinares08} or \cite{lughausen13}.

To include the external field in the $z$-direction, we substitute $g_{z_1}+g_{z_e}$ for $g_{z_1}$ and $g_{z_2}+g_{z_e}$ for $g_{z_2}$, and this affects Eqs~$\ref{eqn:qm1}$-$\ref{eqn:qm6}$. $g_{z_e}$ is the Newtonian value for the external gravitation field.

\end{document}